\documentclass[seceq]{ptptex}
\usepackage[xdvi,dvips]{graphicx}

\newcommand{\bfk}{\mbox{{\boldmath $k$}}}
\newcommand{\bfp}{\mbox{{\boldmath $p$}}}
\newcommand{\bfq}{\mbox{{\boldmath $q$}}}
\newcommand{\bfx}{\mbox{{\boldmath $x$}}}
\newcommand{\bfgamma}{\mbox{{\boldmath $\gamma$}}}
\newcommand{\e}{{\rm e}}
\newcommand{\beq}{\begin{eqnarray}}
\newcommand{\eeq}{\end{eqnarray}}
\newcommand{\btem}{\bibitem}

\newcommand{\Tr}{{\rm Tr}}

\title{%
Pre-critical Phenomena of Two-flavor Color Superconductivity
in Heated Quark Matter
}
\subtitle{
diquark-pair fluctuations and non-Fermi liquid behavior
}

\author{
Masakiyo \textsc{Kitazawa}$^{ab}$\thanks{masky@ruby.scphys.kyoto-u.ac.jp}, 
Tomoi \textsc{Koide}$^c$\thanks{tkoide@yukawa.kyoto-u.ac.jp}, 
Teiji \textsc{Kunihiro}$^a$\thanks{kunihiro@yukawa.kyoto-u.ac.jp} 
and Yukio \textsc{Nemoto}$^d$\thanks{nemoto@quark.phy.bnl.gov}
}

\inst{
$^a$ Yukawa Institute for Theoretical Physics, Kyoto University, Kyoto,
606-8502, Japan,\\
$^b$ Department of Physics, Kyoto University, Kyoto, 606-8502, Japan,\\
$^c$ Institute f\"ur Theoretische Physik, J.W.Goethe Universit\"at, 
D-60054 Frankfurt, Germany,\\
$^d$ RIKEN-BNL Research Center, BNL, Upton, NY 11973
}

\abst{
We investigate  the fluctuations of the diquark-pair field 
and their effects on observables above the
critical temperature $T_c$ in 
 two-flavor color superconductivity (CSC) at moderate density
using a  Nambu-Jona-Lasinio-type effective model of QCD.
Because of the strong-coupling nature of the dynamics, 
the fluctuations of the pair field develop a collective mode, which
has a prominent strength even well above $T_c$.
We show that the collective mode is actually the soft mode
of  CSC.
We examine the effects of the pair fluctuations on  
the specific heat and the quark spectrum for $T$ above but close to $T_c$.
We find that the specific heat 
exhibits  singular behavior because of the pair fluctuations,
in accordance with the general theory of 
second-order phase transitions. 
The quarks display a typical non-Fermi liquid behavior,
owing to the coupling with the soft mode, 
leading to  a pseudo-gap in the density of states of the quarks 
in the vicinity of the critical point.
Some experimental implications of the precursory phenomena 
are also  discussed.
}

\begin{document}

\maketitle

\section{Introduction}

In recent years, there have been many studies of
the properties of dense and cold matter\cite{ref:CSCrev}.
In such a system,
the baryonic density is so high that quarks and gluons
are expected to be deconfined to make a 
quark matter\cite{ref:collins-perry}.
Then, the attractive quark-quark interaction 
in some channels should give rise to a Cooper instability leading to 
color superconductivity (CSC) at sufficiently 
low temperatures\cite{ref:Barr,ref:BL}.
The asymptotic freedom of QCD allows us
to use perturbation theory
to describe CSC at extremely high densities with $\mu > 10^8$ MeV,
where $\mu$ is the quark chemical potential\cite{ref:RS00}.
In this region, the weak-coupling theory analogous to BCS theory, 
i.e. a mean-field approximation (MFA), is valid.
In such a system composed of u, d and s quarks at extremely
high density, it is believed that the quark matter takes the special 
form of the color-superconducting phase, 
i.e. the color-flavor locked (CFL) phase,\cite{ref:cfl}
in which all kinds of quarks  take part equally in the pairing.

What is the phenomenological significance of 
CSC including the CFL?
Possible relevant systems consisting dense QCD matter
include the cores of compact stars,\cite{ref:compact-stars} 
both in equilibrium and in the newly-born stage
and the intermediated states created in
the heavy-ion collisions that are expected to be performed
in the forthcoming facilities at GSI and 
J-PARC\cite{ref:KKKN1,ref:Vosk04}.
One of the basic points concerning  real systems of this kind
lies in the fact 
 that  the quark-number chemical potential $\mu$ and hence the baryon density
are  moderate; $\mu$  should be at most
$\sim$500 MeV.
Therefore the ideal situation realized in an extremely dense system
may  not be expected in these real systems but 
various complications come into play in the determination of 
the nature of CSC.

To realize CSC in a compact star at 
vanishing temperature,
 the color- and electric-charge neutrality conditions must be 
satisfied as well as the $\beta$-equilibrium 
condition.\cite{IB02,ABR00,ref:SRP02}
These conditions, in turn,  induce a mismatch of the Fermi momenta of 
quarks of different flavors and colors.
This is the case in particular when 
the constituent mass of the strange quark,
$M_s$ (which ranges
from some 100 to 500 MeV owing to the dynamical symmetry breaking of
the chiral symmetry) is comparable with $\mu$.
Thus we see that the incorporation of the charge neutrality and
$\beta$-equilibrium conditions, with 
finite $M_s$ taken into account,  brings about quite interesting
complications in the physics of CSC 
in quark matter at moderate densities.
Indeed, it has been shown in the MFA 
that the combination of all these effects can lead to a 
variety of pairing patterns, including the so-called gapless CFL's
\cite{ref:SH03,ref:GLW03,ref:HS03,ref:AKR04,ref:IMTH04,ref:RSR04,ref:FKR04,ref:AKK}.

In this paper, we focus on another important feature of 
QCD matter at moderate densities, i.e.
the strong-coupling nature of QCD at low energy scales, 
which invalidates the MFA.
This strong coupling may imply the significance of large fluctuations of 
the diquark-pair field, especially in the vicinity of the critical 
temperature, $T_c$\cite{ref:KKKN1,ref:Mats00,ref:AHI02}.
It is noteworthy 
that some recent analyses of the RHIC experiments 
 suggest that  quasi-bound quark-antiquark states  
may be formed in moderately {\it hot} QCD matter. 
This reflects the strong-coupling nature of 
QCD\cite{ref:sQGP},
although the possible existence of hadronic modes above $T_c$
was suggested earlier in Refs.~\citen{ref:HK85} and \citen{ref:DeTar85}
\footnote{
See also  recent lattice results presented in Ref.~\citen{ref:Lattice}.
}.

It is thus natural to expect that
 heated quark matter at a moderate density
may also accommodate pre-formed diquark pairs as a pre-critical
phenomenon of CSC phase transition.
Owing to the strong-coupling nature inducing large fluctuations
of the order parameter,
CSC in heated quark matter at moderate densities 
may have some of the same basic properties 
as the superconductivity in strongly correlated electron systems, 
such as high-$T_c$ superconductivity(HTSC), 
rather than usual superconductivity in metals.
\cite{ref:KKKN1,ref:KKKN3} 
We notice that materials at $T>T_c$ show  various types of  
non-Fermi liquid behavior, 
one of which is  {\it pseudogap} formation, 
i.e. an anomalous depression of the density of states (DOS) 
$N(\omega)$ as a function of the fermion energy $\omega$ 
around the Fermi surface\cite{ref:TS99,ref:HTSC1,ref:HTSC2}. 
 It would be intriguing to explore whether
 quarks in the heated quark matter near the critical point 
exhibit  similar abnormal behavior.
In this paper, 
we investigate pair fluctuations of CSC for $T$ above $T_c$
and explore their effects on some physical quantities, 
including the quark spectra in such  quark matter near the critical point.

The relevant quark  matters we have in mind are
those created by  the heavy-ion collisions or 
those realized in the core of a  proto-compact star just after a supernova 
explosion. In these systems,
 the temperature is considerably  high,
and for this reason the $\beta$-equilibrium condition may not be
completely satisfied, in contrast to the case in the 
interiors of cold compact stars.
Therefore, the difference between the chemical potentials of
the up- and down-quarks in these systems
is smaller than that under the $\beta$-equilibrium condition,
and hence a two-flavor superconductor (2SC) will be 
favored  over other pairings 
which incorporate strange quarks.
Furthermore,
the effect of the difference in  chemical potentials induced by
the neutrality conditions will be 
smaller at finite temperature  than that at $T=0$, 
because the Fermi surface is diffused  for $T\not=0$\cite{ref:HS03}.
Thus, one may simply introduce the same chemical potential for 
all the kinds of quarks as a fair approximation to study CSC 
in such hot quark matter.

There are two types of  fluctuations in superconductors, i.e.,
the fermion-pair and gauge-field fluctuations,
irrespective of whether they are 
 electric or color superconductors.
Gauge-field fluctuations are known to make CSC phase
transition  first order 
in the weak coupling region\cite{ref:BL,ref:MIHB04,ref:GHRR,ref:Pisa}.
In this region, it is known that CSC is  strong
type-I superconductivity\cite{ref:MIHB04,ref:GR03}
in which the fluctuations of the gauge field
dominate the pair-field fluctuations.
On the other hand, as is argued in Ref.~\citen{ref:GR03},
CSC is expected to be type-II superconductivity at lower density, i.e.
for $\mu \lesssim 500$MeV,
\footnote{
In Ref.~\citen{ref:GR03}, it is shown
by extrapolation from the weak coupling region
that CSC is type-II superconductivity when $T_c > 14$ MeV 
for $\mu=400$ MeV.
In our model, $T_c = 40$ - $60$ MeV for $\mu=400$ - $500$ MeV,
as  shown in the next section, and 
thus one may conjecture 
that type-II superconductivity is realized.
}
where the fluctuations of the pair field 
dominate  those of the gauge field in contrast to the weak-coupling case.
In the present work, therefore, we simply ignore
the gluon degrees of freedom, and examine the effects of 
pair fluctuations near $T_c$,
as done in Refs.~\citen{ref:KKKN1} and \citen{ref:KKKN3}:
Because pair fluctuations are inherent
in second- (or weak first-) order phase transitions, 
the results in the present work should hold irrespective of 
the different pairing patterns and the chemical potential combination,
as long as the phase transition 
to CSC is second order or  weak first order.
We treat the system at relatively low density, 
where the strange quark degrees of freedom do  not come into play;
accordingly the pairing pattern is taken as 
2SC throughout this paper.

To investigate a system at relatively low temperature and density, 
a perturbative QCD calculation is inadequate,
because of the strong coupling.
Lattice Monte Carlo simulations for finite $\mu$
are still immature for the present purpose, 
although much progress has been being made in recent 
years\cite{ref:finite-density-QCD}.
Therefore, it is appropriate to adopt 
a low-energy effective theory of QCD.
Because we are considering the situation in which
the diquark  pairing dominates over that of the gluons,
an effective model  composed  solely of the quark fields
 may be adopted.
Such effective models include
 the instanton induced model\cite{ref:SS96} and
the Nambu-Jona-Lasinio (NJL) model\cite{ref:HK,ref:Bub04}.
Note that the latter model is a simplified version of the former.
Both models are in fact used to explore the phase structure at low density
\cite{ref:Bub04,ref:BR99,ref:SKP,ref:KKKN2,ref:RSSV}.
One should note here that 
an NJL-type theory can be deduced as a low-energy effective
theory for dense quark matter
on the basis of the renormalization-group equations\cite{ref:EHS99}.
It is also known that the phase structures 
obtained in these models are qualitatively consistent,
and they are similar to those calculated using 
the Schwinger-Dyson equation with the one-gluon exchange 
interaction\cite{ref:AHI02,ref:HT02}.
In this work, we employ the NJL model 
to explore the fluctuations in CSC.

In previous short communications, 
Refs.~\citen{ref:KKKN1} and \citen{ref:KKKN3},
we studied the precursory phenomena to CSC:
We showed for a vanishing wave number that the dynamical fluctuations
of the pair field have a  prominent strength near the vanishing frequency 
up to $T \simeq 1.2T_c$
\cite{ref:KKKN1},
and for this reason that these fluctuations can give rise to 
interesting precursory phenomena, such as 
a pseudogap in the quark DOS near 
$T_c$.\cite{ref:KKKN3}\footnote{
Possible pseudogap formation in association with the 
{\em chiral transition} due to {\it phase} fluctuations
was discussed previously\cite{ref:Ba-Za}.
}
It is worth emphasizing that Ref.~\citen{ref:KKKN3} 
is the first investigation to explore
whether and how the quasi-particle 
properties of quarks are changed 
by the precursory paring mode in CSC.
In the present paper, 
we  present a formulation of the basic theory and
 give a detailed account of both analytic and 
numerical calculational procedures. We  also present a more
extensive study of the properties of the precursory pair field
with finite wave-numbers and
the mechanism through which the pseudogap appears.
We investigate for the first time  the effects of the precursory pair 
fluctuations on the specific heat.
It is found that the heated quark matter close to the critical
point of CSC shows  typical non-Fermi liquid behavior, owing to the 
precursory pair fluctuations, which form a soft mode.

This paper is organized as follows.
In \S 2, we  introduce our model 
 Lagrangian
and present the phase diagram obtained in the MFA in the model.
In \S 3, we  investigate the behavior of the pair fluctuations
above $T_c$ using linear response theory.
It is shown that 
the fluctuations of the pair field develop a collective mode with
 a large strength even well above $T_c$.
We  show that  the complex frequency of  the collective pair-field
moves toward the origin  in the complex energy plane,
which implies that the pair fluctuations form the {\em soft mode} of
CSC phase transition.
We calculate the spectral function of the pair fluctuations,
$\rho(\bfk,\omega)$, as a function of the momentum $\bfk$ and
energy $\omega$ in order to elucidate the spatial and temporal behavior of 
the pair fluctuations when the temperature is lowered toward $T_c$.
We  also present the behavior of the dynamical structure factor.
In \S 4, we  discuss 
the effect of the soft mode on the 
specific heat $c_{\rm v}$ above $T_c$ and show that 
$c_{\rm v}$ increases  when $T$ is lowered to $T_c$ and
eventually diverges at $T=T_c$ in accordance with the singular growth 
of pair fluctuations.
We  examine how the soft mode affects the
 single quark spectrum and changes the quasi-particle picture 
of the fermion above $T_c$ in \S 5.
We  start from a calculation of  the single-quark Green function 
in  the T-matrix approximation to incorporate 
the effects of the pair fluctuations in the quark sector.
It is shown that the pair fluctuations give rise to
a large decay width for  quarks near the Fermi surface, i.e.
 non-Fermi liquid behavior of the quarks near $T_c$.
It is further shown that
the anomalous behavior leads to a pseudogap in the DOS of the quarks.
The chemical potential  dependences 
of the quark spectrum and the DOS are also examined.
The final section is devoted to 
a summary and  concluding remarks.

\section{Model and Phase Diagram}

In this section, after introducing our model Lagrangian, 
we   recapitulate the derivation of the thermodynamic potential
and present the phase diagram obtained in the MFA\cite{ref:KKKN2}. 
This forms the basis for the succeeding investigation of
the nature of the precursory pair fluctuations.

We employ  a Nambu--Jona-Lasinio (NJL) model with  two flavors and
three colors, as mentioned in the Introduction:
\begin{eqnarray}
{\cal L} = \bar{\psi} i/\hspace{-0.2cm}\partial  \psi
+ {\cal L}_S + {\cal L}_C.
\end{eqnarray}
Here the quark-antiquark and quark-quark interactions, 
${\cal L}_S$ and ${\cal L}_C$ are given by
\begin{eqnarray}
{\cal L}_S &=& 
G_S [(\bar{\psi}\psi)^2
+ (\bar{\psi} i\gamma_5 \vec{\tau}\psi)^2],
\nonumber \\
{\cal L}_C &=&
G_C
(\bar{\psi} i\gamma_5 \tau_2 \lambda_A \psi^C)
(\bar{\psi}^C i\gamma_5 \tau_2 \lambda_A \psi),
\label{eqn:Lag}
\end{eqnarray}
with $\psi^C (\bfx) \equiv C\bar{\psi}^T (\bfx)$
and $C = i\gamma_2\gamma_0$ being the charge conjugation operator.
The matrices $\tau_2$ and $\lambda_A$ $(A=2,5,7)$ 
are the antisymmetric components of the Pauli and Gell-Mann matrices 
for the flavor $SU(2)_f$ and color $SU(3)_c$, respectively.
We 
take the chiral limit putting $m_u=m_d=0$, since 
the properties of the diquark condensates are 
affected very little by the small quark masses.

We choose the scalar coupling constant as $G_S = 5.01\, {\rm GeV}^{-2}$ 
and the three-dimensional momentum cutoff $\Lambda = 650$MeV
so as to reproduce the pion decay constant $f_{\pi}=93$ MeV 
and the chiral condensate $\langle \bar{\psi}\psi \rangle=(-250 {\rm MeV})^3$ 
in the chiral limit\cite{ref:Kle}.
There are several sources to determine
 the diquark coupling constant $G_C$,
for example,
the instanton-induced interaction and 
the diquark-quark picture of baryons\cite{ref:AS01}.
The former gives $G_C/G_S=0.5$\cite{ref:RSSV}, 
while the values $G_C/G_S=0.49-0.73$\cite{ref:BT} and
$G_C/G_S=1.5-2$\cite{ref:MBIY} have been obtained using the latter model.
In the present work we fix $G_C/G_S$=0.62, i.e.,
$G_C = 3.11 \, {\rm GeV}^{-2}$, following 
Refs.~\citen{ref:SKP,ref:KKKN1} and \citen{ref:KKKN3}, where $G_C$ is chosen 
so as to reproduce phase diagram similar to that calculated in the
instanton-induced interaction\cite{ref:BR99}.

To determine the phase diagram, we have to first derive 
the thermodynamic potential $ \Omega = -T \log \Tr~ \e^{-\beta K} $
with 
\begin{eqnarray}
K = \int d^3 \bfx \left\{ 
\bar\psi ( -i\bfgamma\cdot\vec\nabla )\psi 
-\mu \bar\psi\gamma^0\psi - {\cal L}_S -{\cal L}_C
\right\}.
\end{eqnarray}
Here, Tr  denotes a trace operation over the color, flavor and
Dirac indices. 
To apply the MFA, we assume a finite
diquark condensate for the 2SC pairing $\Delta$ 
and the quark-antiquark condensate $M$,
\begin{eqnarray}
\Delta = -2 G_C \langle
\bar{\psi}^C \Gamma \psi \rangle, \qquad
M =  - 2 G_S \langle \bar{\psi} \psi \rangle,
\label{eqn:cond}
\end{eqnarray}
with $\Gamma \equiv i\gamma_5 \tau_2 \lambda_2$.
Employing the MFA for ${\cal L}_C$ and ${\cal L}_S$, 
the thermodynamic potential in the MFA per unit volume 
is given by\cite{ref:KKKN2}
\begin{eqnarray}
\Omega_{MF}(M, \Delta ; T,\mu)
&=& -\frac TV \log \Tr \e^{-\beta K}
\nonumber \\
&=& \frac{ M^2 }{ 4G_S } + \frac{ |\Delta|^2 }{ 4G_C }
 -4 \int \frac{ d^3 p }{ (2\pi)^3 }
\left\{ E_p + T \log \left( 1 + \e^{ -\beta \xi_- } \right)
\left( 1 + \e^{ -\beta \xi_+ } \right) \right.\nonumber \\
&&\left. + \epsilon_- + \epsilon_+ + 2T \log
\left( 1 + \e^{ -\beta \epsilon_- } \right)
\left( 1 + \e^{ -\beta \epsilon_+ } \right) \right\},
\label{eqn:Omega}
\end{eqnarray}
where
\begin{eqnarray}
E_p = \sqrt{p^2 + M^2}, \qquad
\xi_{\pm} = E_p \pm \mu, \qquad
\epsilon_{\pm} = \sqrt{\xi^2_{\pm} + |\Delta|^2},
\end{eqnarray}
and $\beta=1/T$.
 The thermodynamic potential 
$\Omega_{MF}$  realizes its absolute minimum as a function of $M$ and $\Delta$
in the equilibrium state; accordingly
the optimal values of  $M$ and $\Delta$ 
 satisfy the stationary conditions
\begin{eqnarray}
\frac{\partial\Omega_{MF}}{\partial M}\biggl|_\Delta = 0 \qquad
{\rm and}\qquad 
\frac{\partial\Omega_{MF}}{\partial\Delta}\biggl|_M = 0,
\label{eqn:GapEq}
\end{eqnarray}
which are actually   self-consistent equations for the condensates
and are called ``the gap equations''.

\begin{figure}[tb]
\begin{center}
\includegraphics[scale=.5]{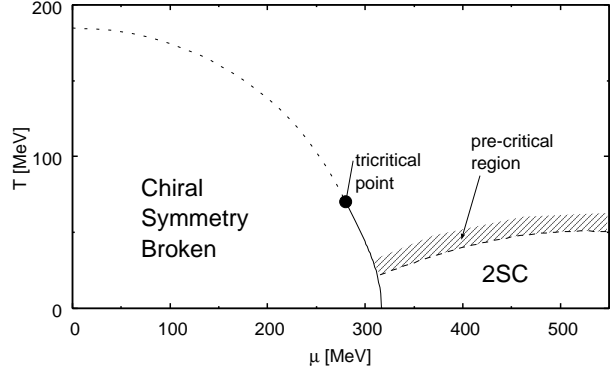}
\caption{
The calculated phase diagram in the $T$-$\mu$ plane for our model.
The solid and dashed curves denote the critical lines 
 of a first- and second-order phase transitions, 
respectively.
}
\label{fig:PhaseDiagram}
\end{center} 
\end{figure}

Using Eqs.~(\ref{eqn:Omega}) and (\ref{eqn:GapEq}), 
we can determine the phase structure of the model.
The phase diagram in the $T$-$\mu$ plane is shown in 
Fig. \ref{fig:PhaseDiagram}:
The dashed (solid) curves
denote the critical line for a second-(first-)order phase transition.
The critical chemical potential
for the chiral-to-CSC transition at $T=0$ is $\mu=316$ MeV
in our model.
The second order normal-to-2SC phase transition occurs
somewhere in the range $T = 35-60$ MeV for $\mu = 350-500$ MeV.
In the following sections, we explore the pair fluctuations and
precursory phenomena above $T_c$ in this region of $\mu$.

For later convenience, we  now explicitly write down the critical
condition for the normal-to-CSC phase transition
with  vanishing chiral condensate ($M=0$).
We first give the explicit expression
of the gap equation  for $\Delta$ in Eq.~(\ref{eqn:GapEq}):
\begin{eqnarray}
\Delta &=& 8 G_C \Delta \int \frac{ d^3 p }{ (2\pi)^3 }
\left\{ \frac1{ \epsilon_- } \tanh \frac{ \beta\epsilon_- }2
+ \frac1{ \epsilon_+ } \tanh \frac{ \beta\epsilon_+ }2 \right\},
\label{eqn:GED}
\end{eqnarray}
where $\epsilon_{\pm}=\sqrt{\vert p\pm\mu\vert ^2+\vert \Delta\vert^2}$,
because we have set $M=0$. 
Dividing Eq.~(\ref{eqn:GED}) by $\Delta$ and then setting
$\Delta=0$, the condition for determining the critical temperature
$T_c$ of the normal-to-CSC phase transition is obtained as
\begin{eqnarray}
0 &=& 1 - 8 G_C \int \frac{ d^3 p }{ (2\pi)^3 }
\left\{ \frac1{ p - \mu } \tanh \frac{ \beta_c ( p - \mu ) }2
+ \frac1{ p + \mu } \tanh \frac{ \beta_c ( p + \mu ) }2 \right\},
\label{eqn:GE_Tc}
\end{eqnarray}
with $\beta_c\equiv 1/T_c$.
Below we find that Eq. (\ref{eqn:GE_Tc}) plays a crucial role for 
understanding the 
anomalous behavior of the pair fluctuations near the critical point.

\section{Pair fluctuations above $T_c$}\label{chap:PF}

In this section, 
we  discuss the properties of the precursory fluctuations
of the diquark-pair field 
in the normal phase on the basis of linear response theory.
Because we limit our attention to the behavior of fluctuations
in the normal phase, we  set $M=\Delta=0$.
It will be shown that a collective mode
 corresponding to pair fluctuations is developed near the
 critical point; we find that this mode is the soft mode for the 
CSC phase transition in the sense  
that  the pole of the spectral function in the
complex energy plane moves toward the origin as $T$ is lowered toward $T_c$.
We then calculate  the spectral function and the structural factor
of the pair field with finite energy and momentum and 
examine their temperature dependence.
We  also discuss how the chemical potential $\mu$
affects the properties of the soft mode.

\subsection{Linear response of the pair field}

In this subsection, we  apply linear response theory in order to
investigate the diquark-pair fluctuations\cite{ref:KKKN1}.
The presentation of the theory here is more formal than that given in
Ref.~\citen{ref:KKKN1}.
See also Ref.~\citen{ref:KKKN1} for a more intuitive discussion
of the linear response of the pair field.

When we apply an external perturbation $H_{ex}(t)$ 
to a thermal equilibrium state at $t=t_0$, 
the expectation value of an arbitrary operator, 
$\langle O( \bfx,t) \rangle$, deviates from the initial equilibrium 
value $\langle O( \bfx,t ) \rangle_{eq}$, where
$O( \bfx,t ) \equiv \e ^{iKt}O( \bfx )\e^{-iKt}$,
 with $K$ being a time-independent
operator.
In the linear response theory, this deviation is given by 
\begin{eqnarray}
\langle O( \bfx,t ) \rangle - \langle O( \bfx,t ) \rangle_{eq}
= i\int^{t}_{t_0} ds \langle [ H_{ex}(s),O( \bfx,t ) ] \rangle _{eq}.
\label{eqn:LRT}
\end{eqnarray}
Here, $\langle~~~\rangle_{eq}$ represents the thermal expectation value 
 without $H_{ex}(t)$.

In order to study the pair fluctuations of CSC,
we choose
\begin{eqnarray}
O( \bfx ) &=& -2G_C \bar{\psi}( \bfx )
\Gamma \psi^C ( \bfx ), \\
H_{ex}(t) &=& \int d^3 \bfx
\Delta^{*}_{ex}( \bfx,t )
\left( \bar{\psi}^C(\bfx) \Gamma \psi(\bfx) \right) + h.c.,
\end{eqnarray}
where $ \Delta^{*}_{ex}( \bfx,t ) $ is a classical external source.
Substituting these into Eq.~(\ref{eqn:LRT})
and using the fact that
$\langle \bar{\psi} \Gamma \psi^C \rangle_{eq}$ 
vanishes in the normal phase, 
we obtain the diquark pair field induced by $\Delta^*_{ex}$,
\begin{eqnarray}
\Delta^{*}_{ind} ( \bfx,t )
&\equiv& -2G_C \langle \bar{\psi}^{C}( \bfx,t)
\Gamma \psi( \bfx,t ) \rangle ,
\nonumber \\
&=& -2G_C\int^{\infty}_{-\infty}dt' \int d^3 {\bfx}'  D^R 
({\bfx},t;{\bfx}',t') \Delta^{*}_{ex}({\bfx}',t').
\label{eqn:DDD}
\end{eqnarray}
Here, the response function $D^R$ is given by
\begin{eqnarray}
D^R ( \bfx,t ; \bfx',t')
&=& -i \theta(t-t')\langle 
[\bar{\psi}({\bfx},t)\Gamma \psi^{C}({\bfx},t), 
\bar{\psi}^{C}({\bfx}',t') \Gamma \psi({\bfx}',t')]
\rangle_{eq} , \nonumber \\
&=& \int \frac{d^3 {\bfk}d\omega}{(2\pi)^4} D^R ( \bfk,\omega )
\e ^{ -i\omega (t-t') }
\e ^{i{\bf k} \cdot ({\bf x}-{\bf x} ')},
\label{eqn:D^R}
\end{eqnarray}
where we have taken the limit $t_0 \to -\infty$.
The Fourier transformation of Eq.~(\ref{eqn:DDD}) gives
\begin{eqnarray}
\Delta^*_{ind} (\bfk,\omega )
=-2G_C D^R ( \bfk,\omega ) \Delta^*_{ex} (\bfk,\omega ).
\label{eqn:DDD_F}
\end{eqnarray}

Equation (\ref{eqn:DDD_F}) contains a significant amount of 
information regarding elementary
excitations in the system. 
We first note that if the external field has the form
\beq
\Delta^*_{ex}(\bfx, t)=
\Delta^*_{ex}(\bfk,\omega )\e^{-i\omega t+i{\bf k}\cdot{\bf x}},
\eeq
then Eq.~(\ref{eqn:DDD_F}) implies that there appears an induced pair  field
\beq
\Delta^*_{ind}(\bfx, t)
=\Delta^*_{ind}(\bfk,\omega )\e^{-i\omega t+i{\bf k}\cdot{\bf x}}
\eeq
with amplitude 
$\Delta^*_{ind}(\bfk,\omega )$.
If the system has an intrinsic
collective excitation with the dispersion relation
$\omega=\omega_s(\bfk)$,
the induced pair field will have a
larger amplitude $\Delta^*_{ind}(\bfk,\omega )$ for
$\omega\sim \omega_s(\bfk)$ owing to a resonance mechanism;
this implies that
$D^R( \bfk,\omega )$ increases as $\omega$ approaches $\omega_s(\bfk)$
on account of Eq.~(\ref{eqn:DDD_F}) and eventually may diverge at  
$\omega=\omega_s(\bfk)$.
Conversely, if the amplitude
$\Delta^*_{ind}(\bfk,\omega )$ has a finite value 
with an  infinitesimally-small external disturbance
$\Delta^*_{ex}(\bfk,\omega )$,
 the system would have an intrinsic collective excitation 
 with frequency $\omega$ and wave number $\bfk$. 
This situation is realized when
the response function   $D^R ( \bfk,\omega )$ is divergent,
which may occur for a special $\omega$ with a given $\bfk$.
Thus we find that the equation
\beq
{D^R ( \bfk,\omega )}^{-1}=0
\label{eqn:coll}
\eeq
gives the dispersion relation $\omega=\omega_s(\bfk)$ 
of the possible collective excitation in the diquark channel
of the system.
A remark is in order. Note that 
 the frequency $\omega_s(\bfk)$ with  given
$\bfk$ can be
a complex number; for instance,
(i) if Re~$\omega_s(\bfk)\not=0$, the mode is 
oscillatory, and  (ii) if $\omega_s(\bfk)$ is pure-imaginary,
the mode is diffusive.
Below we find that
Re~$\omega_s(\bfk)$ is finite, but so small that the 
mode is almost diffusive.

The information concerning the strength of the pair fluctuations is
contained in the spectral function $\rho( \bfk,\omega )$,
which 
 is defined
through the response function $ D^R( \bfk,\omega ) $:
\begin{eqnarray}
\rho ( \bfk,\omega )
= -\frac{1}{\pi } \mbox{Im}D^R ( \bfk,\omega ).
\label{eqn:spectral}
\end{eqnarray}

The strength of the fluctuations is  also 
represented by 
 the dynamical structure factor, defined by
\begin{eqnarray}
S( \bfk,\omega )
&=& -\frac{1}{\pi ( 1-\e^{-\beta \omega} ) }
{\rm Im}~ D^R ( \bfk,\omega ),
\label{eqn:DSF}
\end{eqnarray}
which is positive definite.
At finite temperature,
$S( \bfk,\omega )$ can be directly observed
in scattering experiments\cite{ref:vanHove};
the van Hove scattering formula\cite{ref:vanHove}
gives the cross section in terms of the dynamical structure 
factor.
As we  see below, $\rho( \bfk,\omega )$ represents
the total width of the fluctuations, i.e.,
the difference between the decay rate and the production rate,
and it is negative in the case $\omega<0$ for bosonic excitations.
Accordingly, $S( \bfk,\omega )$, not $\rho( \bfk,\omega )$, represents
the excitation probability of the fluctuations of the system
at finite temperature.

To calculate $D^R( \bfk,\omega )$,
we employ the imaginary time formalism;
we first calculate the response function in the imaginary time formalism
(the two-particle Matsubara Green function),
\begin{eqnarray}
{\cal D}( \bfx,\tau;{\bf0},0 )
&=& - \langle
T_{\tau}
\bar{\psi}( \bfx,\tau)\Gamma \psi^{C}( \bfx,\tau)
\bar{\psi}^C ( {\bf 0},0 ) \Gamma\psi( {\bf 0},0)
\rangle
\nonumber \\
&\equiv&
T\sum_n \int \frac{d^3 \bfk}{(2\pi)^3} 
{\cal D}( \bfk,\nu_n )
\e^{-i\nu_n \tau}\e^{ i{\bf k \cdot x } }, 
\label{eqn:ex-R}
\end{eqnarray}
with 
$\nu_n = 2\pi n/\beta$ being the Matsubara frequency for bosons.

\begin{figure}[tb]
\begin{center}
\includegraphics[scale=.65]{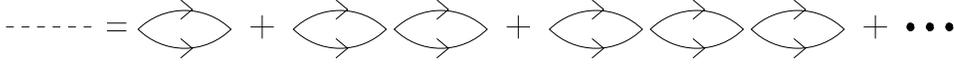} 
\caption{
The diquark propagator in the random phase approximation.
}
\label{fig:RPA}
\end{center} 
\end{figure}

In order to evaluate Eq.~(\ref{eqn:ex-R}), 
we employ the random phase approximation (RPA)  in which
we sum up the ring diagrams shown in Fig.~\ref{fig:RPA},
with the lines 
representing the free Matsubara Green function
in the normal phase ($ M=\Delta=0 $),
\begin{eqnarray}
{\cal G}_0 ( \bfk,\omega_n )
= \frac1 { ( i\omega_n + \mu ) \gamma_0 - \bfk\cdot\bfgamma },
\label{eqn:G^0}
\end{eqnarray}
and the vertices correspond to ${\cal L}_C$.
In this approximation, the Matsubara Green function 
(\ref{eqn:ex-R})  is given by
\begin{eqnarray}
{\cal D}( \bfk,\nu_n )
= \frac12 \frac{{\cal Q}( \bfk,\nu_n )}{1+G_C{\cal Q}( \bfk,\nu_n )},
\label{eqn:D}
\end{eqnarray}
where ${\cal Q} (\bfk,\nu_n)$ is the contribution from the 
one-loop particle-particle correlation function,
\begin{eqnarray}
{\cal Q}( \bfk,\nu_n ) 
= -2 T \sum_m \int \frac{ d^3 \bfp }{ (2\pi)^3 } \mbox{Tr}
\left[
C \Gamma {\cal G}_0( \bfk-\bfp ,\nu_n-\omega_m )
\Gamma C {\cal G}_0^T( \bfp ,\omega_m )
\right]. \nonumber \\
\label{eqn:Q}
\end{eqnarray}
Here $\omega_n = (2n+1)\pi/\beta $ denotes the Matsubara frequency
for fermions. 
The response function in the real time Eq.~(\ref{eqn:D^R}) is 
given by the analytic continuation $ i\nu_n \to \omega+i\eta$,
and we obtain
\begin{eqnarray}
\left. D^R (\bfk,\omega ) 
= {\cal D} ( \bfk,\nu_n ) \right|_{i\nu_n=\omega + i\eta}
= \frac{1}{2}\frac{Q^R( \bfk,\omega ) }{ 1 + G_C Q^R( \bfk,\omega ) },
\label{eqn:D^R_RPA}
\end{eqnarray}
with $Q^R( \bfk,\omega ) 
= {\cal Q} ( \bfk,\nu_n ) |_{i\nu_n=\omega + i\eta} $.
It is thus seen that the equation to determine the dispersion relation for
the collective excitation given in Eq. (\ref{eqn:coll}) is reduced to
\beq
1 + G_C Q^R( \bfk,\omega )=0.
\label{eqn:coll-2}
\eeq

\begin{figure}[tb]
\begin{center}
\includegraphics[width=12cm]{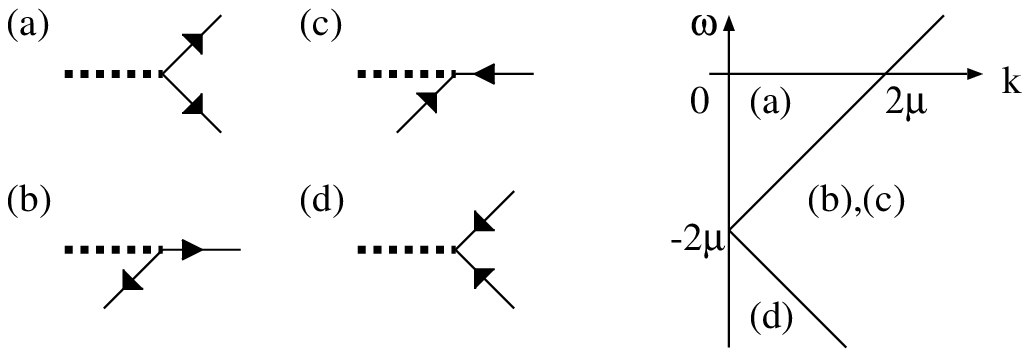}
\caption{ 
The left panel shows the kinetic processes contained in
${\rm Im} Q^R( \bfk,\omega )$; 
the corresponding inverse processes are not shown.
The dashed line denotes the diquark pair field and the solid
line the quark or the antiquark.
The right panel represents the energy-momentum regions in which 
the decay processes (a)-(d) (and their inverse ones) occur.
} 
\label{fig:decay}
\end{center}
\end{figure}

With some manipulations 
(see Appendix \ref{sec:appQ} for details),
we find the following simplified form of $Q^R( \bfk,\omega )$:
\begin{eqnarray}
Q^R( \bfk,\omega) 
&=& N_f ( N_c-1 )
\int \frac{ d^3 \bfp }{ (2\pi)^3 } \frac 1{ e_1 e_2 }
\nonumber \\
& \times& \left\{
\left[ ( e_1+e_2 )^2 - k^2 \right]
\left(
\frac{ 1- f^- ( e_2 ) - f^-( e_1 ) }
{ \omega + 2\mu - e_1 - e_2 +i\eta } +
\frac{ 1- f^+ ( e_2 ) - f^+( e_1 ) }
{ \omega + 2\mu + e_1 + e_2 +i\eta }  
\right) \right. \nonumber \\
&-& \left. 
\left[ ( e_1-e_2 )^2 - k^2 \right]
\left(  
\frac{ f^- ( e_2 ) - f^+( e_1 ) }
{ \omega + 2\mu + e_1 -e_2  +i\eta }  +
\frac{ f^+ ( e_2 ) - f^- ( e_1 ) }
{ \omega + 2\mu - e_1 + e_2 +i\eta }
\right) \right\}. \nonumber \\
\label{eqn:Q_explicit}
\end{eqnarray}
Here
$f^{\pm}(x)=1/[\e^{\beta(x\pm \mu)}+1]$ is the Fermi-Dirac distribution
function for the quark and anti-quark, respectively, and
 $e_1 \equiv |\bfp|$, $e_2 \equiv |\bfk-\bfp|$ 
and $k \equiv |\bfk|$.
\footnote{
Using the change of variables $\bfp \leftrightarrow \bfk-\bfp$,
Eq.~(\ref{eqn:Q_explicit}) is converted into $Q^R( \bfk,\omega )$ 
of Ref.~\citen{ref:KKKN1}.
Equation~(\ref{eqn:Q_explicit}) is more convenient for the following
calculation than that given in Ref.~\citen{ref:KKKN1}.
}
The imaginary part of $Q^R( \bfk,\omega )$ is given by
\begin{eqnarray}
\mbox{Im} Q^R( \bfk,\omega )
&=&
-\pi N_f ( N_c-1 ) \int \frac{d^3\bfp}{ (2\pi)^3 }
\frac{ (\omega+2\mu)^2 - k^2 }{ e_1 e_2 } \nonumber \\
&\times& \left[
\left( 1 - f^-( e_1 ) - f^-( e_2 ) \right) \right.
\delta( \omega + 2\mu - e_1 - e_2 ) \nonumber \\
&&
- \left( 1 - f^+( e_1 ) - f^+( e_2 ) \right)
\delta( \omega + 2\mu + e_1 + e_2 ) \nonumber \\
&&
+ \left( f^-( e_1 ) - f^+( e_2 ) \right)
\delta( \omega + 2\mu - e_1 + e_2 ) \nonumber \\
&&
\left.
- \left( f^+( e_1 ) - f^-( e_2 ) \right)
\delta( \omega + 2\mu + e_1 - e_2 ) \right].
\label{eqn:ImQ_explicit}
\end{eqnarray}
Each term in  the square brackets of Eq.~(\ref{eqn:ImQ_explicit}) 
corresponds to the decay process shown 
in Figs.~\ref{fig:decay} (a)-(d) and its inverse.
To see this, the following identities for 
arbitrary quantities $f_i$ 
($i=1, 2$) are useful:
$1- f_1-f_2= (1-f_1)(1-f_2) -f_1f_2$ and
$f_1-f_2=(1-f_2)f_1 -f_2(1-f_1)$. 
In general,
 the imaginary part of the retarded Green function is related
 to the net decay rate or the {\em total} width.
Because of the delta function in each term
owing to energy-momentum conservation, 
the ranges of  energy and momentum over which each decay process 
can take place are restricted to the region shown 
in the right panel of Fig.~\ref{fig:decay};
for example, the decay process 
into two particles, as shown in (a) of Fig.~{\ref{fig:decay}}, and
 its inverse occur only for $ \omega > |\bfk|-2\mu $.
The processes (b) and (c),  the so-called Landau damping processes,
and their inverses take place only for $ -|\bfk|-2\mu < \omega <
|\bfk|-2\mu $;
hence the collective modes with vanishing momenta do not decay through 
these processes.
We now see that the decay process (a) plays the
dominant role 
near $T_c$, because the strength of the pair fluctuations 
is concentrated near $ (\omega,\bfk)=(0,{\bf0}) $ for $T\sim T_c $,
as we  show in the next subsection.
We note that $D^R(\bfk,\omega)$ is analytic around 
$ (\omega,\bfk) = (0,{\bf0}) $ for $T>T_c$, 
because the functions $Q^R( \bfk,\omega )$,
 and hence $D^R( \bfk,\omega )$,
have discontinuities only  on the boundaries
$\omega = \pm|\bfk|-2\mu$.
\footnote{
In the case of  particle-hole modes,
a discontinuity of the response function exists 
at $\omega = \pm|\bfk|$\cite{ref:HK}.
Hence, the response function is not analytic at the origin.
}

Here, we would like to discuss the critical behavior
of the response function
Eq.~(\ref{eqn:D^R_RPA}).
In general, when the temperature approaches $T_c$ of the 
second-order transition from above,
the fluctuations of the order parameter with a low frequency 
(small $\omega$)
and a long wave-length (small $\bfk$) become easily excited.
This implies,  on account of Eq.~(\ref{eqn:DDD_F}), 
that $D^R( \bfk,\omega )$ with  small $\omega$ and $\bfk$ 
becomes larger as the system approaches the critical point.
At  $T_c$,
the system becomes unstable with respect to {\em uniform}
 diquark-pair formation, and a finite and permanent pair-field is formed
with an infinitesimally-small external field.
Then, according to Eq.(\ref{eqn:DDD_F}),
$D^R( {\bf 0}, 0 )$ should be divergent at $T=T_c$.
One can show that this is indeed the case, that is
\beq
{D^R({\bf 0},0)}^{-1}\Big\vert_{T=T_c}=0.
\label{eqn:thouless-0}
\eeq
 In fact,
the denominator of Eq.~(\ref{eqn:D^R_RPA}) 
for vanishing $\bfk$ and $\omega$ becomes
\begin{eqnarray}
\lefteqn{  1 + G_C Q^R( {\bf 0},0 )  }\nonumber \\
&=&
1 - 8 G_C \int \frac{ d^3 \bfp }{ (2\pi)^3 }
\left\{ \frac1{ p-\mu } \tanh \frac{ \beta( p-\mu ) }2
+ \frac1{ p+\mu } \tanh \frac{ \beta( p+\mu ) }2
\right\}, \nonumber \\
\label{eqn:Thouless}
\end{eqnarray}
which is  found to vanish at $T=T_c$, on account of 
Eq.~(\ref{eqn:GE_Tc}).
Equation (\ref{eqn:thouless-0}) 
is called the Thouless criterion\cite{ref:Thou},
which may be used to determine the critical point.

For the momentum integration of Eq.~(\ref{eqn:Q_explicit}), 
we employ the  following cutoff scheme\cite{ref:NJL61,ref:KLW}.
First, we  note that the imaginary part of $Q^R( \bfk,\omega )$
is free from an ultraviolet divergence.
Thus, we can evaluate the imaginary part without introducing a cutoff.
In this scheme, the imaginary part, Eq.~(\ref{eqn:ImQ_explicit}), 
can be reduced to a compact form (see Appendix~\ref{sec:appQ}),
\begin{eqnarray}
{\rm Im}~Q^R ( \bfk,\omega )
&=&-N_f (N_c-1)T \frac{(\omega+2\mu)^2 -k^2}{2\pi k}
\nonumber \\
&&\times \left[
\log \frac{\cosh (\omega+k)/4T}{\cosh (\omega-k)/4T}
- \theta(-|\omega+2\mu|+k)\frac{\omega}{2T}
\right].
\label{eqn:ImQ}
\end{eqnarray}
Then, we evaluate the real part of $Q^R( \bfk,\omega )$ 
from the imaginary part by using the dispersion relation.
We introduce a cutoff at this stage,
because ${\rm Re}~Q^R( \bfk,\omega )$ has an ultraviolet divergence.
The cutoff should be chosen 
so as to satisfy the Thouless criterion for $k=0$,
discussed above.
Thus, as shown in Appendix~\ref{sec:appQ}, the real part of $Q^R$ is
expressed as  
\begin{eqnarray}
{\rm Re}Q^R (\bfk,\omega)
= -\frac 1\pi {\rm P}
\int^{2\Lambda -2\mu}_{-2\Lambda -2\mu} 
d\omega' \frac{ {\rm Im} Q^R (\bfk,\omega')}{ \omega-\omega' },
\label{eqn:DispRel}
\end{eqnarray}
where P denotes the principal value.\footnote{
The Thouless criterion does not restrict
the range of integration in Eq.~(\ref{eqn:DispRel}) for $ k \ne 0$,
because this criterion is a property of static homogeneous matter.
In this sense, we have no criterion to determine 
the range of integration in Eq.~(\ref{eqn:DispRel}) for $ k\ne 0 $.
In this work, however, we simply assume that the range 
does not change even for finite momentum.
In any case, the final results for the relevant range of $|\bfk|$ and $\omega$
do not depend on the choice of cutoff scheme.
}
This cutoff scheme has the advantages
 that  the imaginary part of $Q^R( \bfk,\omega )$
does not destroy any conservation laws through the 
introduction of a cutoff and
it does reflects the symmetries of the system.
In addition, the  simple form Eq.~(\ref{eqn:ImQ}), 
which is derived without introducing a cutoff, is quite convenient
for numerical calculations.

\subsection{Collective excitation}

In this subsection, we  obtain the 
dispersion relation 
of the collective mode in the diquark channel
 by solving Eq.~(\ref{eqn:coll-2}) for $\omega$, with $\bfk$ given.
Solutions of Eq.~(\ref{eqn:coll-2}) should exist
in the lower-half complex energy plane $\mathbb{C}^-$, 
because the imaginary part of the pole should be negative;
otherwise the system would be  unstable with respect of the creation of 
collective modes.
Thus we actually solve the equation,
\begin{eqnarray}
1 + G_C Q^R ( \bfk, z ) = 0 
\label{eqn:poleQ}
\end{eqnarray}
 for a complex variable $\omega\equiv z$ 
with $ Q^R( \bfk,z ) \equiv Q^R( \bfk,\omega)|_{ \omega \to z }$.
In fact, it can be numerically verified that 
Eq.~(\ref{eqn:poleQ}) does not have any solution 
in the upper-half plane $\mathbb{C}^+$.
In order to calculate Eq.~(\ref{eqn:poleQ}) for $z\in\mathbb{C}^-$,
we must perform the analytic continuation of $Q^R(\bfk,\omega)$
to $\mathbb{C}^-$;
a simple analytic continuation of Eq.~(\ref{eqn:Q}),
defined as
\beq
 Q( \bfk,z ) \equiv {\cal Q}( \bfk,\nu_n)|_{ i\nu_n \to z },
\qquad z\in\mathbb{C},
\eeq
with $\mathbb{C}$ denoting the entire complex plane,
has a cut along the real axis. Hence 
$Q^R(\bfk,z) \ne Q(\bfk,z)$ in $\mathbb{C}^-$.
The retarded function $Q^R(\bfk,z)$ in $\mathbb{C^-}$ 
is defined on  another Riemann sheet of $Q(\bfk,z)$, and therefore
one has to derive the analytic function of $Q(\bfk,z)$
in this Riemann sheet to solve Eq.~(\ref{eqn:poleQ}).
As shown in Appendix~\ref{sec:appQ2},
the form of $Q^R(\bfk,z)$ in $\mathbb{C}^-$ is found to be
\begin{eqnarray}
Q^R ( \bfk,z ) = Q ( \bfk,z ) + 2iI ( \bfk,z ),
\label{eqn:QR_lower}
\end{eqnarray}
with $I( \bfk,z ) \equiv \mbox{Im} Q^R( \bfk, \omega )|_{\omega\to z}$.

\begin{figure}[tb]
\begin{center}
\includegraphics[scale=.8]{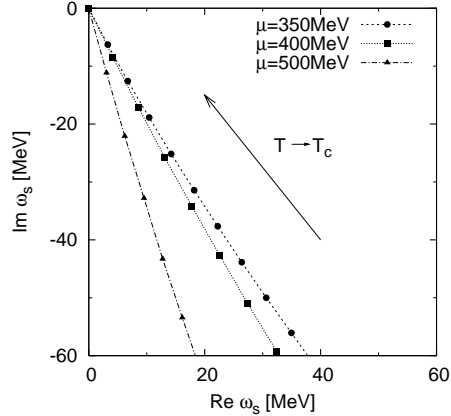} 
\caption{
The pole $\omega_s$ of the collective mode above $T_c$
for vanishing momentum with various $\mu$.
The points denote the positions of the poles at the reduced temperature 
$\varepsilon \equiv (T-T_c)/T_c = 0$, $0.1$, $0.2\cdots$.
Note here that $\omega_s (k=0) \vert_{\varepsilon =0}=0$.
The pole approaches the origin as $\varepsilon$ is lowered.
}
\label{fig:pole}
\end{center} 
\end{figure}

Using Eq.~(\ref{eqn:QR_lower}), we find a solution 
of Eq.~(\ref{eqn:poleQ}) in $\mathbb{C}^-$.
Note that the Thouless criterion Eq.(\ref{eqn:Thouless}) ensures
that there  exists a solution at the origin for $T=T_c$,
irrespective of $\mu$. The temperature dependence of the solutions with $k=0$
is  already presented in Ref.~\citen{ref:KKKN1}.
We display them  in Fig.~\ref{fig:pole} for three chemical potentials, 
$\mu=350$, $400$ and $500$ MeV; the three curves correspond
to these three values of $\mu$, while
the dots on the curves denote the position of the poles
for each reduced temperature,
 $\varepsilon \equiv (T-T_c)/T_c= 0, 0.1, 0.2,\cdots$.
It  can be seen that the pole goes up to the origin in the complex
energy plane as $T$ is lowered toward $T_c$.
It is also seen that 
the rate at which the origin is approached is almost the same for all the 
chemical potentials.
The result clearly shows that the collective mode of the pair
fluctuations is the soft mode\cite{ref:HK} of CSC phase transition,
which gives rise to the precursory phenomena of CSC
phase transition, as  shown in the following sections.
We will see
that the softening  of the pole causes a rapid growth of the peak
in $\rho(\bfk,\omega)$ and $S(\bfk,\omega)$ as $T\to T_c$.

It is  noteworthy that ${\rm Re}~\omega$ 
is small but finite.
In the weak coupling limit, as described by BCS theory, 
the pole of the soft mode
appears on the imaginary axis, i.e., ${\rm Re}~\omega=0$. 
This is due to a particle-hole symmetry 
inherent in the weak coupling limit:
In metal superconductors, 
an attractive interaction exists only for fermions with 
energies satisfying
 $\vert \omega \vert <\omega_D$, as  measured from the Fermi energy $E_F$  
with the Debye frequency $\omega_D$, which
satisfies $\omega_D \ll E_F$.
Thus, the DOS for the fermions participating in the pairing 
can be treated  as a constant in this energy range, and  
accordingly the range
is symmetric with respect to the Fermi surface.
For CSC,
the pairing involves all of the states 
in the Fermi sphere, not just those around the Fermi surface.
This leads to an asymmetry  between particles and holes.
This asymmetry is the origin of the finite real part of the soft mode.
From Fig.~\ref{fig:pole}, we find that the ratio
${\rm Re}~\omega/ {\rm Im}~\omega$ becomes smaller for larger $\mu$.
This is plausible, because the rate of the particle-hole asymmetry 
becomes smaller as $\mu$ is increased.
It is instructive to note that 
an increase of ${\rm Im}~\omega$, along  with that of $\mu$, 
is also understood in a different way:
The density of states
on the Fermi surface into which the collective mode decays 
increases with the chemical potential.
Thus, a larger decay probability of the
collective mode is realized, and this leads to a larger ${\rm Im}~\omega$.

The fact that the pole has both real and imaginary parts implies that
the dynamical behavior of the order parameter near $T_c$ is
a damped-oscillator mode, while the pure imaginary poles
in the weak coupling limit correspond to over-damped modes, or
diffusive modes as mentioned above.
However, the absolute value of ${\rm Im}~\omega$ is larger 
than that of  ${\rm Re}~\omega$ even in the present case.
This implies that the dynamical behavior of 
the order parameter of CSC is approximately the same as that of 
the over-damped case,
and it is accurately described by a non-linear diffusion equation like 
the time-dependent Ginzburg-Landau (TDGL) equation
in the weak coupling theory\cite{ref:Cyr,ref:KKKN1}.

\begin{figure}[tb]
\begin{center}
\includegraphics[scale=1]{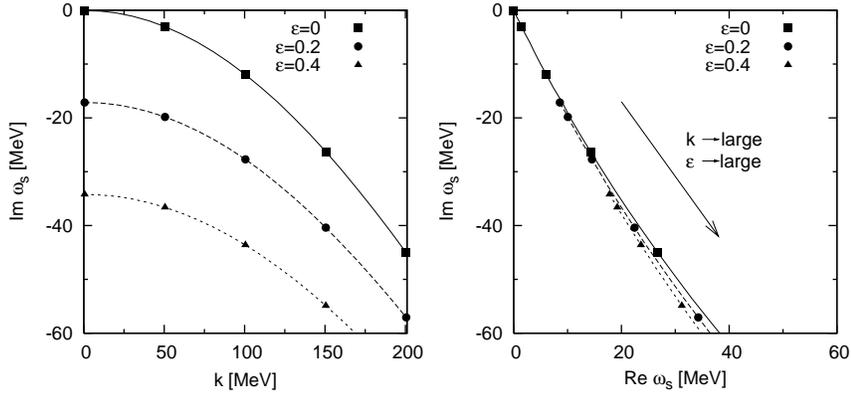} 
\caption{
Momentum dependence of the pole $\omega_s(k)$ for $\mu=400$ MeV
and $\varepsilon=0$, $0.2$ and $0.4$.
The left panel depicts the $k$ dependence of Im$\omega_s (k)$.
The right panel depicts the momentum dependence of $\omega_s (k)$
in the complex energy plane.
The pole moves away from the origin as $k$ is increased.
}
\label{fig:pole_fk}
\end{center} 
\end{figure}

Next, let us consider the momentum dependence of the pole $\omega_s(k)$.
In Fig.~\ref{fig:pole_fk},
we plot ${\rm Im}\omega_s(k)$
as a function of $k$.
It is seen that $\left| {\rm Im}\omega_s(k) \right|$ becomes larger 
as $k$ is increased.
This means that the lifetime of the fluctuations of the pair field 
 becomes shorter with  smaller wavelength.
The pole $\omega_s(k)$ in the complex energy plane is shown in the right panel
of Fig.~\ref{fig:pole_fk}. It can be seen that 
the ratio ${\rm Re}\omega/{\rm Im}\omega$ is almost independent of
 $k$ and $\varepsilon$.

\subsection{Softening of the pair fluctuations}

\begin{figure}[t]
\begin{center}\leavevmode
\includegraphics[width=14cm]{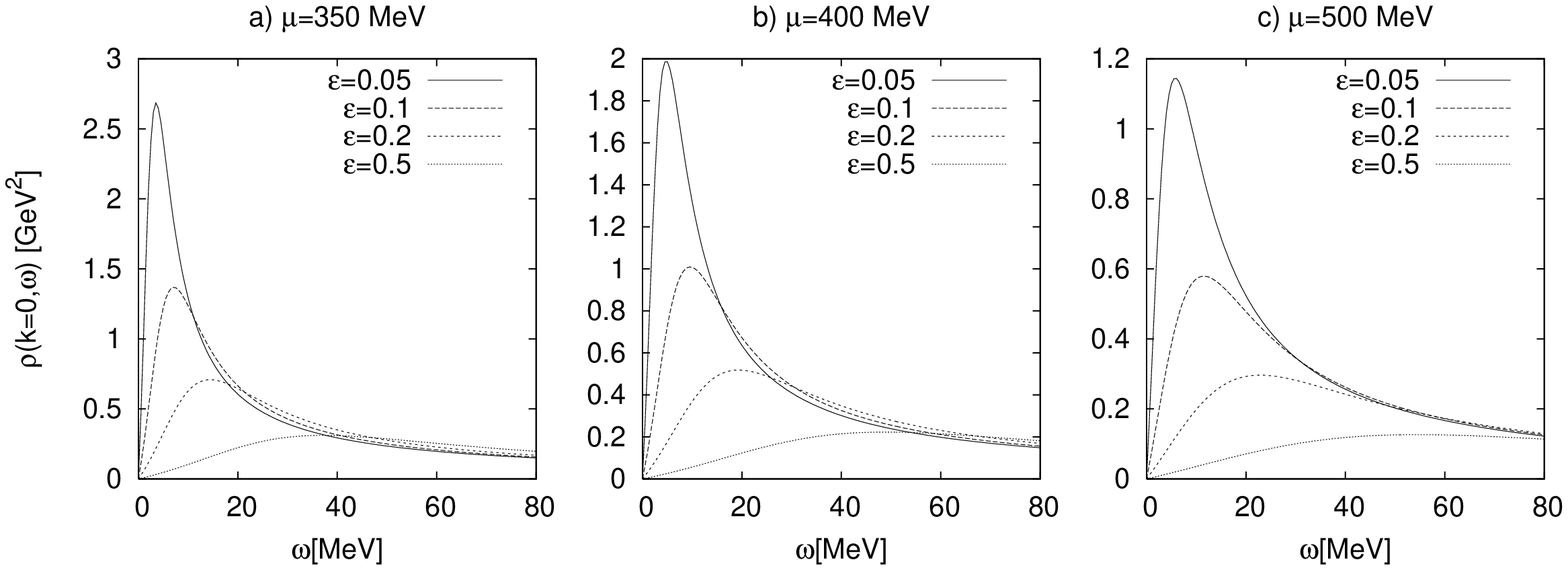} 
\caption{
The spectral function $\rho(\bfk,\omega)$ with $k=0$ 
for $\mu=350, 400$ and $500$ MeV 
and several values of $\varepsilon$.
The height of the peak increases as $T$ is lowered in the normal phase
for all values of $\mu$.
A clear peak is seen even at  as high a temperature as  $\varepsilon = 0.2$.
}
\label{fig:spc1}
\end{center} 
\end{figure}

In the previous subsection, we saw that
quark matter at values of $T$ above but near $T_c$ develops a
collective mode owing to diquark-pair fluctuations.
In this subsection, we  show how the strength of the fluctuations
changes when $T$ is lowered toward $T_c$.

We first show the temperature dependence of $ \rho ( \bfk,\omega ) $
at vanishing momentum $\bfk=\bf0$ for three different chemical potentials, 
$\mu = 350,400,500$ MeV, and at several reduced temperatures
in Fig. \ref{fig:spc1}\cite{ref:KKKN1}.
One can see that the position of the  peak moves 
toward the origin, while the peak height increases as the temperature 
is lowered toward $T_c$, 
although the rate of growth of the peak 
decreases as $\mu$ is increased.

Using the Thouless criterion Eq.~(\ref{eqn:Thouless}),
it can be analytically shown that
the spectral function has a divergent peak at the origin,
$(\bfk,\omega) = ({\bf 0},0)$, for $T=T_c$.
It can be seen in Fig.~\ref{fig:spc1} that the peak remains quite distinct
as away from $T_c$ as $\varepsilon = 0.2$.
In the case of electric superconductivity in metals, 
the effect of fluctuations is small, because of weak coupling.
Precursory phenomena in electric conductivity can be observed 
experimentally. For example,
there appears anomalous enhancement called `paraconductivity'
in the electric conductivity just above the critical temperature.
However, the range of $T$ over which clear paraconductivity is seen
is limited to those satisfying $\varepsilon\lesssim 10^{-3}$,
 even in superconductors
with large fluctuations such as dirty alloys and low-dimensional
materials.
Therefore, we can conclude that fluctuations in CSC survive
for values of $\varepsilon$ in two or three orders larger 
than in the case of  electric superconductors.
It follows that the precursory fluctuation phenomena of CSC can
exist over a rather wide range of $T$, and hence 
they can be used as  experimental signatures of CSC.

\begin{figure}[t]
\begin{center}\leavevmode
(a) \hspace{6.2cm} (b) 
\includegraphics[width=14cm]{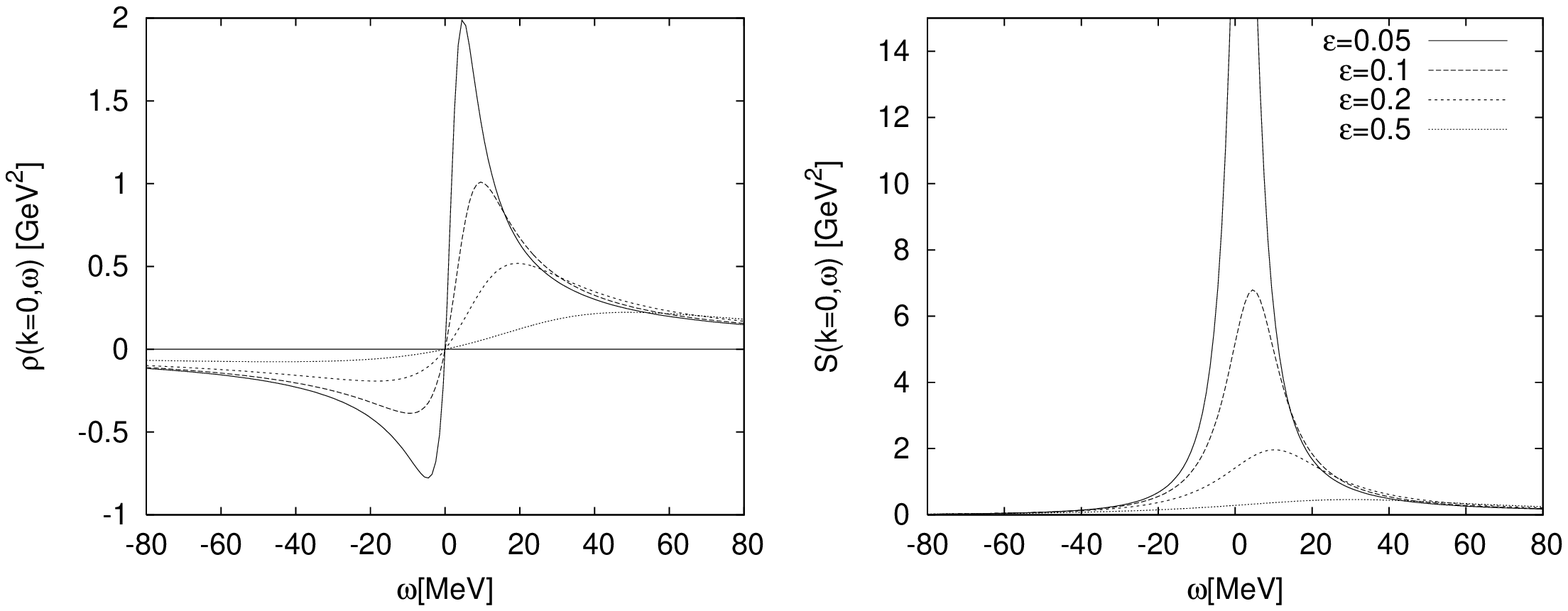} 
\caption{
The spectral function $\rho(\bfk,\omega)$ and 
the dynamical structure factor $S(\bfk,\omega)$ 
for  $k=0$ at $\mu=400$ MeV.
}
\label{fig:spc2}
\end{center} 
\end{figure}

The above arguments are based on the behavior of the spectral function 
$\rho(\bfk,\omega)$.
We can draw further conclusions  on the strength of the fluctuations if 
we consider the dynamical structure factor 
$S( \bfk,\omega )$, defined in Eq.~(\ref{eqn:DSF}).
As mentioned above, $S( \bfk,\omega )$ is more directly related to 
observables than the spectral function $\rho( \bfk,\omega )$, which  
has a negative value for bosons when $\omega<0$,
as shown in Fig.~\ref{fig:spc2}.
We display the $\varepsilon$ dependence of $S( \bfk,\omega )$ 
for $\mu=400$ MeV at vanishing momentum transfer in Fig.~\ref{fig:spc2} (b).
It can be seen that a peak appears in $S(\bfk,\omega)$ and 
moves toward the origin as the temperature approaches $T_c$.
This behavior is consistent with that of the spectral function, and 
it is a reflection of the softening of the pair fluctuations.
We remark that the position of the pole corresponds to the peak of 
the dynamical structure factor, not the spectral function.

\begin{figure}[t]
\begin{center}\leavevmode
\hspace{-1.4cm}
\includegraphics[scale=.65]{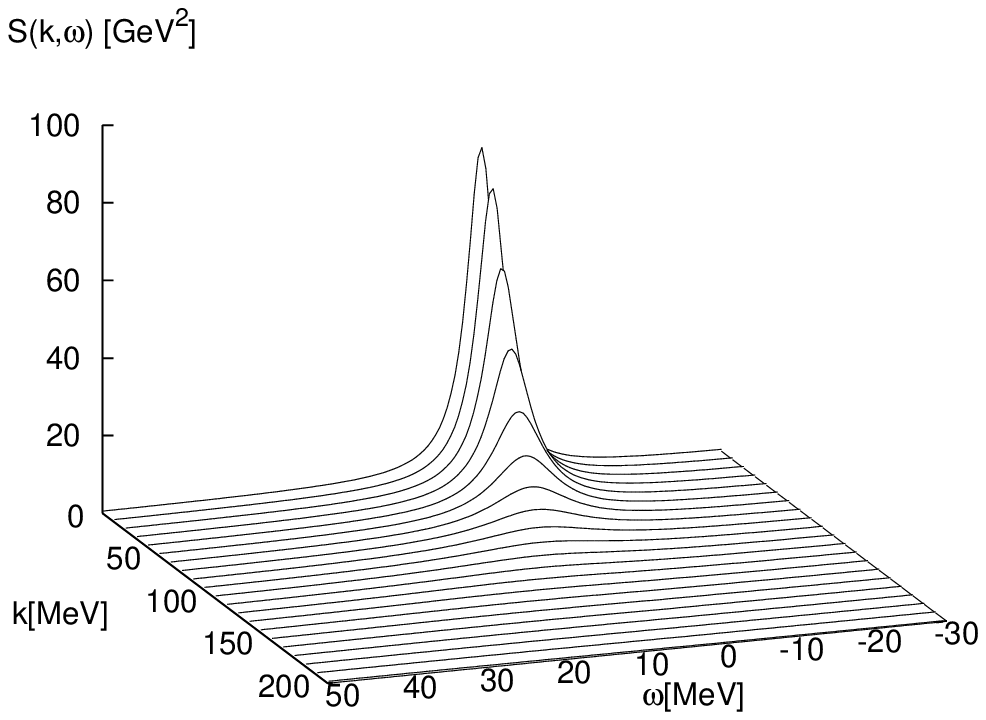} \hspace{-1.8cm}
\includegraphics[scale=.65]{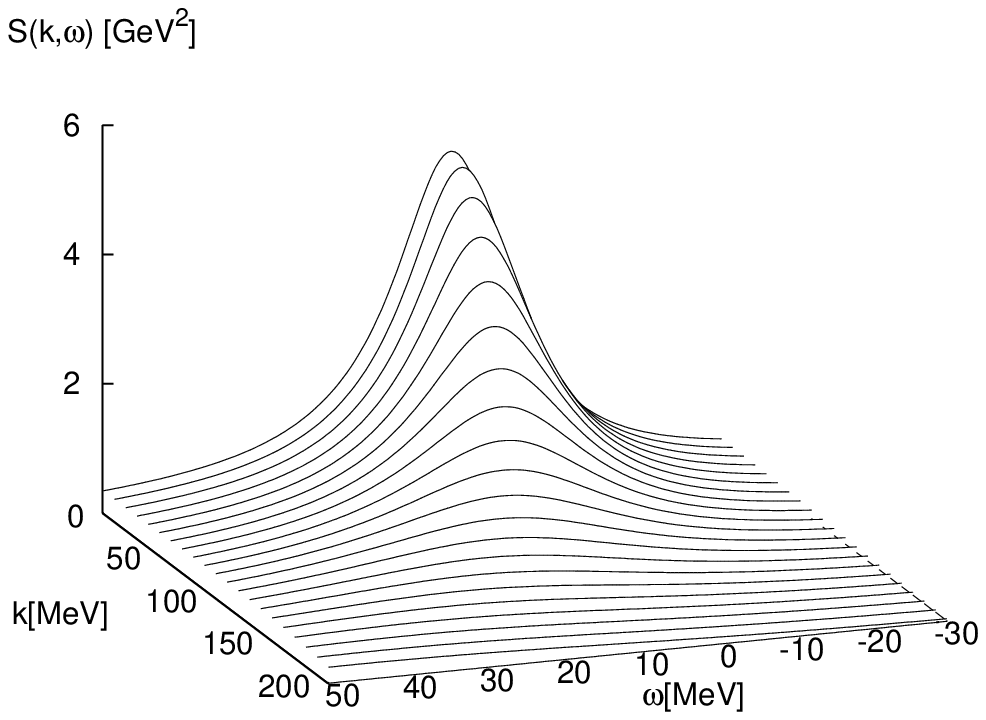} 
\hspace{-1.8cm}
\caption{
The energy and momentum dependence of $S(\bfk,\omega)$ 
for  $\mu=400$ MeV.
Here, we have 
$\varepsilon =0.02$ ($\varepsilon=0.1$) in the left (right) panel.
The peak in $S(\bfk,\omega)$ grows, which implies that the pair fluctuations 
around $\omega=k=0$ become stronger, as $\varepsilon$ approaches $0$.
}
\label{fig:dsf_fk}
\end{center} 
\end{figure}

Next,
we show the energy-momentum dependence of $S( \bfk,\omega) $
at $\mu=400$ MeV and $\varepsilon=0.02$ and $0.1$ in Fig.~\ref{fig:dsf_fk}.
The peak near $\omega=0$ flattens as $k$ is increased
in each figure, which means that 
the fluctuations with smaller wavelength tend to be  suppressed.

\section{Specific heat} \label{chap:SH}

In the previous section, 
we showed that there exist strong pair fluctuations
even well above $T_c$.
In the present and following sections, we  evaluate the effects of 
the pair fluctuations on some observables.
In this section, we calculate the specific heat,
 taking into account the precursory fluctuations above $T_c$, 
and show that the pair fluctuations cause 
an anomalous enhancement of the specific heat
over range of $T$ that is similar to that over which 
a prominent peak in the spectral function is seen.

The specific heat per unit volume, $c_{\rm v}$,  
is calculated from
the thermodynamic potential per unit volume, $\Omega$, as
\begin{eqnarray}
c_{\rm v} = -T \frac{ \partial^2 \Omega }{ \partial T^2 }.
\label{eqn:sh_def}
\end{eqnarray}
Therefore, we  first evaluate the thermodynamic potential $\Omega$
in  the normal phase,
 incorporating the pair fluctuations:
\beq
\Omega = \Omega_0 + \Omega_{\rm fl.}.
\eeq
Here, $\Omega_0$ ($\Omega_{\rm fl.}$) 
denotes the contribution  at the mean-field level (from the
pair fluctuations).
Then, the specific heat $c_{\rm v}$ is also divided into the two
parts,
corresponding to $\Omega_0$ and $\Omega_{\rm fl.}$, respectively:
\beq
c_{\rm v} = c_{\rm v}^0 + c_{\rm v}^{\rm fl.}.
\eeq

\begin{figure}[bt]
\begin{center}
\includegraphics[]{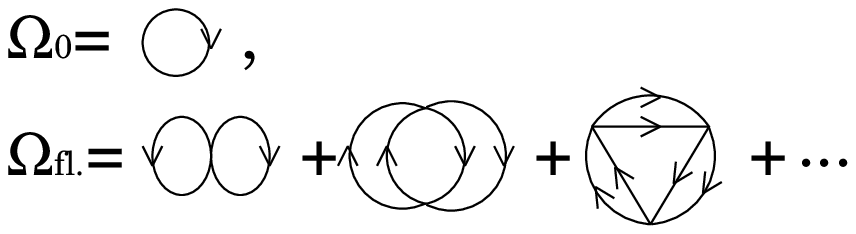} 
\caption{
The diagrams taken into account in the calculation 
of the thermodynamic potential 
$\Omega = \Omega_0 + \Omega_{\rm fl.}$.
Here, $\Omega_0$ is the thermodynamic potential of a free fermionic
system, and $\Omega_{\rm fl.}$ is the contribution 
from the fluctuations of the pair field.
}
\label{fig:Omega}
\end{center} 
\end{figure}

\begin{figure}[bt]
\begin{center}
\includegraphics[scale=.75]{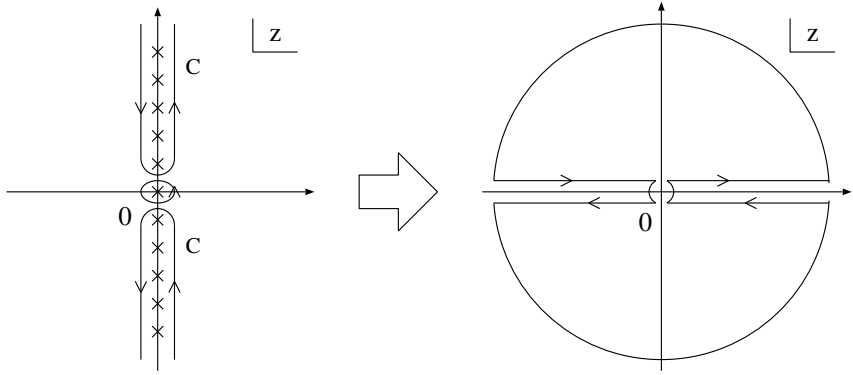} 
\caption{
The contour of integration in Eq.~(\ref{eqn:Omega_fl}).
}
\label{fig:contour_cv}
\end{center} 
\end{figure}

The thermodynamic potential of the free quarks $\Omega_0$ reads
\begin{eqnarray}
\Omega_0 
&=& \Omega_{MF}(M=0,\Delta=0) \nonumber \\ 
&=&
 T \sum_n \int \frac{ d^3\bfk }{ (2\pi)^3 }
{\rm Tr} \log {\cal G}_0( \bfk ,\nu_n ).
\label{eqn:Omega_free}
\end{eqnarray}
Here, $\Omega_{\rm fl.}$ consists of the summation of 
the ring diagrams shown in Fig.~\ref{fig:Omega},
where all vertices correspond to the diquark interaction
term Eq.~(\ref{eqn:Lag}). Below, we find that 
these diagrams correspond to those considered in the response function
$D^R(\bfk,\omega)$ in the RPA,
and hence $\Omega_{\rm fl.}$ exhibits anomalous behavior
through $D^R(\bfk,\omega)$.
The lowest-order diagram in $\Omega_{\rm fl.}$ is calculated to be
\begin{eqnarray}
\Omega_{\rm fl.}^{(1)}
&=& 
-2 G_C T^2 \sum_{m,n} \int \frac{ d^3\bfp_1 d^2\bfp_2 }{ (2\pi)^6 }
\sum_{A=2,5,7} \mbox{Tr} \left[ 
i\gamma_5 \tau_2 \lambda_A {\cal G}( \bfp_1,\omega_m )
i\gamma_5 \tau_2 \lambda_A {\cal G}( \bfp_2,\omega_n ) \right]
\nonumber \\
&=&
-2 G_C T^2 \sum_{m,n} \int \frac{ d^3\bfp_1 d^2\bfp_2 }{ (2\pi)^6 }
3 \mbox{Tr} \left[ 
i\gamma_5 \tau_2 \lambda_2 {\cal G}( \bfp_1,\omega_m )
i\gamma_5 \tau_2 \lambda_2 {\cal G}( \bfp_2,\omega_n ) \right]
\nonumber \\
&=&
3 T\sum_n \int \frac{ d^3\bfp }{ (2\pi)^3 } G_C {\cal Q}( \bfp,\nu_n ).
\label{eqn:Omega_fl1}
\end{eqnarray}
In the second equality here, 
we have used the fact that
the $A=2,5$ and 7 terms in the first line of Eq.~(\ref{eqn:Omega_fl1})
give the same contribution;
we have incorporated them all into the $A=2$ term
with an overall factor of $3$.
Physically, this factor corresponds to the existence of 
three degenerate collective excitations 
of the pair field in the normal phase, that is, 
the {\it red}-{\it blue}, {\it blue}-{\it green} and 
{\it green}-{\it red} collective modes in the color space.
This factor appears in the diagrams of $\Omega_{\rm fl.}$
at all orders.
Similarly,
the $l$-th order diagrams of $\Omega_{\rm fl.}$ can be evaluated as
\begin{eqnarray}
\Omega_{\rm fl.}^{(l)}
= -\frac 3l T \sum_n \int \frac{ d^3\bfp }{ (2\pi)^3 }
\left[ -G_C {\cal Q}( \bfp,\nu_n ) \right]^l.
\end{eqnarray}
Summing up all the terms, we obtain
\begin{eqnarray}
\Omega_{\rm fl.}
= 
\sum_{l=1}^{\infty} \Omega_{\rm fl.}^{(l)} 
&=&
-3 T \sum_n \int \frac{ d^3\bfp }{ (2\pi)^3 }
\sum_{l=1}^{\infty} \frac 1l \left[ -G_C {\cal Q}( \bfp,\nu_n ) \right]^l
\label{eqn:Omega_fl_n} \\
&=& 
3 T \sum_n \int \frac{ d^3\bfp }{ (2\pi)^3 }
\log \left[ 1 + G_C {\cal Q}( \bfp,\nu_n ) \right]
\nonumber \\
&=&
3 \int \frac{ d^3\bfp }{ (2\pi)^3 } \int_C \frac{dz}{2\pi i}
\frac1{ \e^{\beta z} -1 }
\log \left[ 1 + G_C Q( \bfp,z ) \right]
\nonumber \\
&=&
\frac 3\pi \int \frac{ d^3\bfp }{ (2\pi)^3 } {\rm P} \int d\omega
\frac1{ \e^{\beta\omega} -1 }
\mbox{Im} \log \left[ 1 + G_C Q^R( \bfp,\omega ) \right].
\label{eqn:Omega_fl}
\end{eqnarray}
In the last equality, 
the contour of integration is modified so as to avoid the cut
along the real axis, 
as shown in Fig.~\ref{fig:contour_cv}\cite{ref:AGD}.
Note that the argument of the logarithmic function
is the same as  the denominator of the 
response function of the pair field $D^R( \bfk,\omega )$,
which, we have seen, exhibits a singular behavior near $T_c$. 
Therefore, $\Omega_{\rm fl.}$ should also
show an anomalous behavior near $T_c$,
which gives rise to an enhancement of $c_{\rm v}^{\rm fl.}$
\cite{ref:Thou}.

Before taking the derivatives in Eq.~(\ref{eqn:sh_def}),
we expand the argument of the logarithmic function 
in Eq.~(\ref{eqn:Omega_fl}) 
in a Taylor expansion about $\omega=k=0$ and $T=T_c$. To lowest order,
this yields
\begin{eqnarray}
G_C^{-1} + Q^R( \bfk,\omega ) \simeq C_0 \omega + C|\bfk|^2 + A \varepsilon,
\label{eqn:TDGL}
\end{eqnarray}
with
\begin{eqnarray}
C_0 = \left. \frac{ \partial Q^R({\bf 0},0) }{ \partial \omega } 
\right|_{T=T_c},\,
C = \left. \frac{ \partial Q^R({\bf 0},0) }{ \partial |{\bf k}|^2 } 
\right|_{T=T_c},\, 
A = \left. T\frac{ \partial Q^R({\bf 0},0) }{ \partial T } 
\right|_{T=T_c}.
\label{eqn:TDGL_coeff}
\end{eqnarray}
Note that there is no zeroth-order term 
in the expansion Eq.~(\ref{eqn:TDGL}).
This follows from the Thouless criterion, 
$ G_C^{-1} + Q^R( {\bf 0},0 )|_{T=T_c} = 0 $.
One can numerically check that
the right-hand side of Eq.~(\ref{eqn:TDGL}) indeed does provide
a good approximation of
the behavior of the left hand side over a rather wide region 
of $\omega$, $\bfk$ and $\varepsilon$; roughly for those values satisfying
$|\omega|,|\bfk| \lesssim 120$ MeV and 
$\varepsilon \lesssim 0.3$\cite{ref:KK}.
As was shown in the previous section, 
the strength of the pair field is concentrated 
around $ \omega=|\bfk|=0 $ as $T$ approaches $T_c$.
Therefore, the fluctuating modes in the vicinity of the origin
of the $\omega$-$k$ plane 
are expected to give the dominant contribution to
the  thermodynamic potential and hence the specific heat.
Thus, use of the simple expansion given in Eq.~(\ref{eqn:TDGL}) 
in  their calculation is justified.

\begin{figure}[tb]
\begin{center}
\includegraphics[scale=1.1]{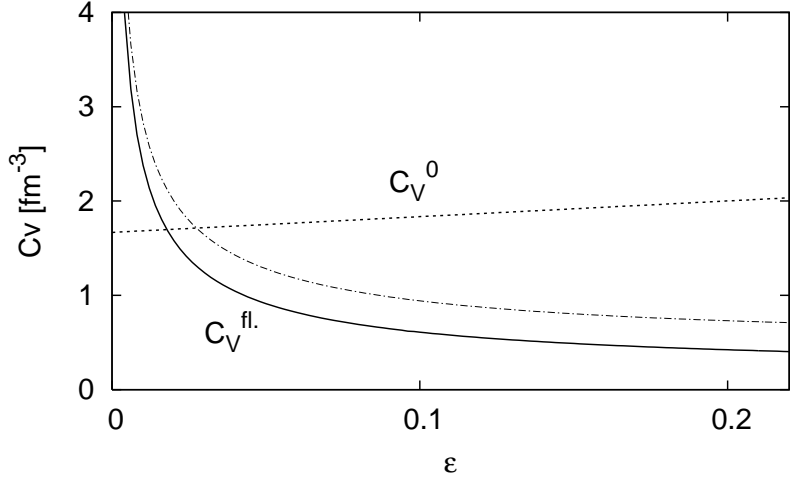} 
\caption{
The specific heat per unit volume 
in the free fermionic system, $c_{\rm v}^0$, and 
the effects of the fluctuations of the pair field,
 $c_{\rm v}^{\rm fl.}$.
The total specific heat of the system is 
$c_{\rm v} = c_{\rm v}^0 + c_{\rm v}^{\rm fl.}$.
It is  seen that as $\varepsilon$ decreases,
the enhancement of $c_{\rm v}^{\rm fl.}$ begins to
become appreciable near $\varepsilon = 0.05$-$0.1$.
The dot-dashed curve represents the specific heat obtained with 
only the static part of $\Omega_{\rm fl.}$.
}
\label{fig:SH}
\end{center} 
\end{figure}

In Fig.~\ref{fig:SH}, 
we plot the behavior of the specific heat
$c_{\rm v}^0$ and $c_{\rm v}^{\rm fl.}$, above $T_c$.
It is seen that $c_{\rm v}^{\rm fl.}$ diverges as $ T \to T_c $.
A clear enhancement of $c_{\rm v}^{\rm fl.}$ is seen 
already near $\varepsilon = 0.05$-$0.1$.
The range of temperatures in which 
$c_{\rm v}^{\rm fl.}$ is larger than $c_{\rm v}^0$
is called the Ginzburg-Levanyuk region\cite{ref:Leva59,ref:Ginz60}.
In our case, the Ginzburg-Levanyuk region exists 
up to $\varepsilon = 0.02$ above $T_c$.
In electric superconductors, the Ginzburg-Levanyuk region
is continued only to values $\varepsilon$
satisfying $\varepsilon \lesssim 10^{-3}$, even for dirty alloys.
Therefore, the Ginzburg-Levanyuk region for CSC
is very much larger than that for  electric superconductors.
It can be numerically checked that 
$c_{\rm v}^{\rm fl.} \sim \varepsilon^{-1/2}$ near $T_c$.
This critical exponent of $c_{\rm v}$
is the same  as that obtained 
from the Ginzburg-Landau equation 
without nonlinear terms\cite{ref:LL}.
This is because we have adopted the RPA 
for the pair field.
However, the RPA is not valid in the vicinity of
$T_c$, 
because there the amplitude of the fluctuations becomes large, 
and therefore the nonlinear effects begin to play a significant role.
For this reason, the true critical exponent of the specific heat 
should differ from $-1/2$.

The singular behavior of the 
specific heat above $T_c$ of the superconductivity  is 
usually studied  using only 
the static part of $\Omega_{\rm fl.}$\cite{ref:LL}, 
i.e., the $n=0$ component of the Matsubara sum
in Eq.~(\ref{eqn:Omega_fl_n}).
However, we have
explicitly included the dynamical modes as well.
We now give an account of the relation between the two 
approaches.
The most singular behavior of $\Omega_{\rm fl.}$
which causes the enhancement of $c_{\rm v}^{\rm fl.}$ near $T_c$
certainly comes from 
the static part of $\Omega_{\rm fl.}$,
because only the static part in Eq.~(\ref{eqn:Omega_fl_n})
diverges as $\varepsilon \to 0$. 
Therefore, the contribution from the static part dominates
the behavior of $\Omega_{\rm fl.}$, at least near $T_c$.
In Fig.~\ref{fig:SH},
the specific heat obtained using only the static part 
is plotted by the dot-dashed curve.
It is seen that the singular behavior of 
$c_{\rm v}^{\rm fl.}$ near $T_c$ is reproduced by 
the contribution from the static part of $\Omega_{\rm fl.}$ alone,
and the contribution from the $n\ne 0$ components of $\Omega_{\rm fl.}$ 
to $c_{\rm v}^{\rm fl.}$ is smaller than $c_{\rm v}^0$.

Recently, Voskresensky calculated the specific heat above the CFL phase
and found that the Ginzburg-Levanyuk region exists
up to $\varepsilon \sim 1$\cite{ref:Vosk04}, which is more than 
one order of magnitude larger in the units of $\varepsilon$ 
than that found in the present case.
The origin of this difference can be understood as follows.
In two-flavor quark matter, 
the degeneracy of the collective modes
gives rise to the factor of $3$ in Eq.~(\ref{eqn:Omega_fl}),
while the existence of nine collective modes
above $T_c$ for the CFL phase can cause a factor of $9$ 
to appear in $c_{\rm v}^{\rm fl.}$,
provided that the strange quark mass is ignored.
Therefore, $c_{\rm v}^{\rm fl.}$ above the CFL phase
can take a value three times larger  than that in ours.
The coefficients of the Ginzburg-Landau equation 
in Ref.~\citen{ref:Vosk04}, which are determined
in the weak coupling limit, further increase $c_{\rm v}^{\rm fl.}$.
These factors 
can account for the wide Ginzburg-Levanyuk region obtained
in Ref.~\citen{ref:Vosk04}.

\section{Non-Fermi liquid behavior due to fluctuations}

As seen in the previous sections,
the pair fluctuations form a well-developed collective mode 
near $T_c$ in CSC,
and they cause anomalous behavior in the specific heat.
In this section, we explore how  the pair fluctuations in turn
affect the properties of quarks.
We  show that the pair fluctuations give rise to 
 non-Fermi liquid behavior of the quarks near $T_c$:
It is shown, for instance, that 
a large decay width is acquired by the quarks near the Fermi surface.
It is also shown that
the anomalous behavior leads to the pseudogap in the DOS of the quarks.

The pair field may exhibit both {\em amplitude} and  {\em phase}
 fluctuations.
We assume that the {\em amplitude} fluctuations dominate 
the {\it phase} fluctuations\cite{ref:HTSC1}. 
Thus we are led to employ the 
T-matrix approximation\cite{ref:KadBay}, which is suitable to
evaluate the effects of the amplitude fluctuations of the pair field
on the quarks.
Historically, the DOS above $T_c$ was first calculated
within the T-matrix approximation 
for the case of the electric  superconductivity
in the weak coupling regime 
more than three decades ago
\cite{ref:Maki68,ref:Taka70,ref:ARW70}.
It was shown that 
a pseudogap can form but only in the vicinity of $T_c$. 
After the discovery of the HTSC,
the same approximation was reconsidered in the strong coupling regime
in the study of the DOS for HTSC\cite{ref:JML97,ref:HTSC2}.
In this section, 
we  compute the quark Green function 
in the T-matrix approximation
and evaluate the spectral function and the dispersion
relation of the quarks in the relativistic kinematics.

\subsection{T-matrix approach}

The one-particle quark spectral function is defined by
\begin{eqnarray}
{\cal A}( \bfk,\omega )
&=&-\frac1\pi \cdot{\rm Im}G^R ( \bfk,\omega )
\nonumber \\
&=& -\frac1\pi \frac{ G^R( \bfk,\omega ) - G^A( \bfk,\omega )}{2i}
= -\frac1\pi \frac{ G^R( \bfk,\omega ) 
- \gamma^0 G^{R\dag}( \bfk,\omega ) \gamma^0 }{2i},
\end{eqnarray}
where $G^R( \bfk,\omega )$ and $G^A( \bfk,\omega )$ are 
the retarded and advanced Green functions of the quarks,
respectively. 
Conversely,
$G^R( \bfk,\omega )$ is given in terms of 
 ${\cal A}( \bfk,\omega )$ as
\begin{eqnarray}
G^R ( \bfk,\omega )
&=& \int d\omega' 
\frac{ {\cal A}( \bfk,\omega' ) }{ \omega-\omega'+i\eta }.
\label{eqn:SR-RGF}
\end{eqnarray}
The spectral function ${\cal A}( \bfk,\omega )$ has 
a Dirac matrix structure in the relativistic formalism.
From the rotational and parity invariances, 
the Dirac indices of ${\cal A}( \bfk,\omega )$ can be decomposed 
into three parts,
\begin{eqnarray}
{\cal A}( \bfk,\omega )=
\rho_0( \bfk,\omega ) \gamma^0 
- \rho_{\rm v}( \bfk,\omega ) \hat{\bfk}\cdot{\bfgamma} +
\rho_{\rm s}( \bfk,\omega ),
\end{eqnarray}
with $\hat{\bfk} = {\bfk}/|{\bfk}|$.
Actually,  $\rho_{\rm s}( \bfk,\omega )$ 
vanishes in the chiral limit, which we have taken. 
Note that $\rho_0( \bfk,\omega )$ is related to the number density
of quarks, and hence to the DOS of the quarks, as 
\begin{eqnarray}
N(\omega) &=& 4\int \frac{d^3 {\bfk}}{(2\pi)^3}
{\rm Tr}_{\rm c,f}\left[ \rho_{0}({\bfk},\omega) \right],
\label{eqn:N_0}
\end{eqnarray}
with ${\rm Tr}_{\rm c,f}$ 
denoting the trace over color and flavor indices.

\begin{figure}[tb]
\begin{center}
\includegraphics{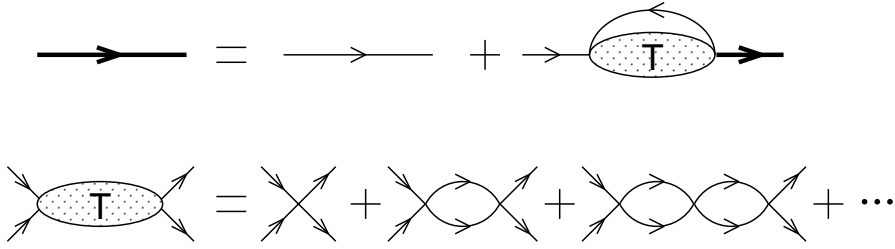} 
\caption{
The Feynman diagrams representing the
quark Green function in the non-self-consistent T-matrix approximation.
The thin lines represent the free propagator ${\cal G}_0$,
while the bold lines represent the full propagator ${\cal G}$.
}
\label{fig:self}
\end{center} 
\end{figure}

To calculate the quark Green function Eq.~(\ref{eqn:SR-RGF}),
we use the Matsubara formalism.
The Dyson-Schwinger equation for the quark Matsubara Green function 
${\cal G}( \bfk,\omega_n )$ reads
\begin{eqnarray}
{\cal G}( \bfk, \omega_n )
&=& {\cal G}_0 ( \bfk, \omega_n ) + {\cal G}_0( \bfk,\omega_n )
\tilde\Sigma( \bfk,\omega_n ){\cal G}( \bfk,\omega_n ), 
\label{eqn:DS-Eq}
\end{eqnarray}
where $ \tilde\Sigma( \bfk,\omega_n )$ is the quark self-energy
in the imaginary-time formalism.
We take the following approximate form for
the self-energy $\tilde\Sigma( \bfk,\omega_n )$ 
\begin{eqnarray}
\tilde\Sigma( \bfp,\omega_n ) 
&=& 4 \sum_{ A=2,5,7 } ( \lambda_A )^2
T\sum_m \int \frac{d^3 \bfk }{(2\pi)^3}
\tilde\Xi ( \bfp+\bfk, \omega_n+\omega'_m )
{\cal G}_0 ( \bfk,\omega'_m ) 
\label{eqn:SE0} \\
&=& 8
T\sum_m \int \frac{d^3 \bfk }{(2\pi)^3}
\tilde\Xi ( \bfp+\bfk, \omega_n+\omega'_m )
{\cal G}_0 ( \bfk,\omega'_m ),
\label{eqn:SE1}
\end{eqnarray}
where the T-matrix $\tilde\Xi(\bfk,\nu_n)$ is
\begin{eqnarray}
\tilde{\Xi} ( \bfk,\nu_n )
= -G_C \frac 1{ 1+G_C {\cal Q}({\bfk},\nu_n) }
= -G_C \left( 1 -2G_C {\cal D}( \bfk,\nu_n ) \right),
\label{eqn:Xi_2}
\end{eqnarray}
with ${\cal D}( \bfk,\nu_n )$ defined in Eq.~(\ref{eqn:D}).
This is the so-called non-self-consistent T-matrix approximation,
which is consistent with
 RPA employed in \S 3 to describe the pair fluctuations.
Figure \ref{fig:self} is the schematic representation of our T-matrix 
approximation; the thin and bold lines represent the free and full
Matsubara propagators, ${\cal G}_0( \bfk,\omega )$ 
and ${\cal G}( \bfk,\omega )$.
If the thin lines in the diagram
are replaced by the bold lines, the approximation becomes
the {\em self-consistent} approximation. 
We  comment on this approximation in the concluding remarks.
The T-matrix $\tilde\Xi( \bfk,\nu_n )$ show
the same anomalous behavior as the response function 
near $T_c$, in accordance with the softening of the pair fluctuations.

In the second equality of Eq.~(\ref{eqn:SE1}), 
we have used the fact that the sum of the Gell-Mann matrices 
in Eq.~(\ref{eqn:SE0}) yields $\sum_A ( \lambda_A )^2 = 2$.
This factor corresponds to the two possible patterns of 
$\tilde\Sigma( \bfk,\omega_n )$ in the color space 
owing to the existence of three degenerate collective modes
of the pair field in the normal phase;
a {\it red} quark, for example, can create 
a {\it red}-{\it green} and {\it red}-{\it blue} collective modes 
incorporating {\it green} or {\it blue} quarks, respectively.
In the case of the specific heat discussed in \S 4, 
the factor $ 3  (=N_c) $ appears
because the three degenerate collective modes contribute equally.
On the other hand, only two of the three collective modes 
can couple to an incoming quark, and this is the reason that 
the factor $ 2 (=N_c-1) $ 
arises in the self-energy appearing in Eq.~(\ref{eqn:SE1}).

The analytic continuation of $\tilde\Sigma( \bfk,\omega_n )$
to the real axis from the upper-half complex energy plane
gives the self-energy in real time,
\begin{eqnarray}
\Sigma^R( \bfk,\omega )
= \tilde\Sigma( \bfk,\omega_n )|_{ i\omega_n \to \omega+i\eta }.
\label{eqn:Sig^R_def}
\end{eqnarray}
After several manipulations (summarized in Appendix~\ref{sec:appSigma}),
this self-energy is found to be
\begin{eqnarray}
\Sigma^R ( \bfp,\omega )
&=&
2 \int \frac{ d^4q}{ (2\pi)^4 }
\left\{
\tanh \frac{q^0}{ 2T }
\Xi^R ( \bfp+\bfq,\omega +q^0 ) {\rm Im} G_0^R( \bfq,q^0 )
\right. \nonumber \\ && \left. \qquad \qquad 
+ \coth \frac{q^0}{2T}
{\rm Im}\Xi^R ( \bfp+\bfq,q^0 ) G_0^A( \bfq,q^0-\omega )
\right\},
\label{eqn:Sig^R}
\end{eqnarray}
with
\begin{eqnarray}
\Xi^R( \bfk,\omega ) 
= \tilde\Xi( \bfk,\nu_n )|_{i\nu_n = \omega + i\eta}
= -[ G_C^{-1} + Q^R( \bfk,\omega ) ]^{-1}.
\end{eqnarray}
The imaginary part of Eq.~(\ref{eqn:Sig^R}) is
\begin{eqnarray}
\lefteqn{ \mbox{Im} \Sigma^R( \bfp,\omega ) }
\nonumber \\
&=&
2 \int \frac{ d^4 q }{ (2\pi)^4 }
\left( \tanh \frac{ q^0 }{2T} - \coth \frac{ q^0 + \omega }{2T} \right)
\mbox{Im} \Xi^R ( \bfp+\bfq, \omega+q^0 )
\mbox{Im} G^R( \bfq,q^0 ). \qquad
\label{eqn:ImSig^R}
\end{eqnarray}
Using Eq.~(\ref{eqn:Sig^R}),
the retarded Green function, Eq.~(\ref{eqn:SR-RGF}), can be written
\begin{eqnarray}
G^R( \bfk,\omega )
= \frac1{  [ G_0^R( \bfk,\omega ) ]^{-1} - \Sigma^R( \bfk,\omega ) }.
\label{eqn:G^R_quark}
\end{eqnarray}

From the rotational and parity invariances, 
we find that the self-energy has the same Dirac matrix structure 
as the spectral function,
\begin{eqnarray}
\Sigma^R ( \bfk,\omega )
= \Sigma_0 ( \bfk,\omega ) \gamma^0 
- \Sigma_{\rm v}( \bfk,\omega ) \hat{\bfk}\cdot\bfgamma
+ \Sigma_{\rm s}( \bfk,\omega ),
\end{eqnarray}
with
$ \Sigma_0 = (1/4)\Tr[ \Sigma^R \gamma^0 ] $,
$ \Sigma_{\rm v} = (1/4)\Tr[ \Sigma^R \hat{\bfk} \cdot \bfgamma ] $
and $ \Sigma_{\rm s} = (1/4)\Tr[ \Sigma^R ] $.
From Eq.~(\ref{eqn:Sig^R}), one can easily check that 
$\Sigma_{\rm s}$ vanishes in the chiral limit, 
which also means that $\rho_{\rm s}=0$ in this case.

It is useful to decompose $\Sigma^R( \bfk,\omega )$ and 
${\cal A}( \bfk,\omega )$ into positive and negative
energy parts using the projection operators
$\Lambda_\mp(\bfk) = ( 1 \pm \gamma^0 \bfgamma\cdot\hat{\bfk})/2$.
We obtain
\begin{eqnarray}
\Sigma^R( \bfk,\omega )
&=& \gamma^0(\Sigma_-( \bfk,\omega ) \Lambda_-(\bfk) 
+ \Sigma_+( \bfk,\omega ) \Lambda_+(\bfk) ),
\nonumber \\
{\cal A}( \bfk,\omega )
&=& (\rho_-( \bfk,\omega ) \Lambda_-(\bfk) 
+ \rho_+( \bfk,\omega ) \Lambda_+(\bfk) ) \gamma^0,
\end{eqnarray}
where
\begin{eqnarray}
\Sigma_{\mp}( \bfk,\omega )
&=& \frac12 \Tr [ \Sigma^R( \bfk,\omega ) \Lambda_\mp(\bfk) \gamma^0 ]
= \Sigma_0( \bfk,\omega ) \mp \Sigma_{\rm v}( \bfk,\omega ),
\label{eqn:Sigma_mp} \\
\rho_\mp( \bfk,\omega ) 
&=& \frac12 \Tr [ {\cal A}( \bfk,\omega ) \gamma^0 \Lambda_\mp( \bfk ) ]
= \frac1\pi {\rm Im} 
\frac1{ \omega +\mu \mp k -\Sigma_\mp( \bfk,\omega ) }
\nonumber \\
&=& \rho_0( \bfk,\omega ) \pm \rho_{\rm v}( \bfk,\omega )
\end{eqnarray}
are the self-energies and spectral functions for
positive and negative energies, respectively.
Using Eq.~(\ref{eqn:Sigma_mp}),
the retarded Green function is written as  
\begin{eqnarray}
G^R ( \bfk,\omega )
&=& 
\frac{ \Lambda_-(\bfk) \gamma^0 }
{ \omega +\mu -k - \Sigma^-( \bfk,\omega ) +i\eta }
+ \frac{ \Lambda_+(\bfk) \gamma^0 }
{ \omega +\mu +k - \Sigma^+( \bfk,\omega ) +i\eta }
\label{eqn:G3} \\
&=&
\frac{ \Lambda_-(\bfk) \gamma^0 }{ R_-(\bfk,\omega ) +i\eta }
+ \frac{ \Lambda_+(\bfk) \gamma^0 }{ R_+(\bfk,\omega ) +i\eta }.
\end{eqnarray}
In the second equality, we have defined
\begin{eqnarray}
R_\mp( \bfk,\omega ) 
\equiv \omega +\mu \mp k - \Sigma_\mp( \bfk,\omega ),
\end{eqnarray}
for later convenience.
The equations
\beq
 {\rm Re}~R_{\mp}( \bfk,\omega_{\mp}) = 0,
\label{eqn:q-disp}
\eeq
give the dispersion relations for  quarks and
anti-quarks $ \omega= \omega_\mp( \bfk ) $, respectively.

There are several possible choices of the four-momentum cutoff
for the integral in Eq.~(\ref{eqn:Sig^R}).
To compute this integral,
we first calculate the imaginary part Eq.~(\ref{eqn:ImSig^R});
substituting the explicit formula for ${\rm Im}~G_0^R( \bfk,\omega )$
to Eq.~(\ref{eqn:ImSig^R}), the $q^0$ integral is removed
by the delta function in ${\rm Im}~G_0^R( \bfk,\omega )$
(see, Eqs.~(\ref{eqn:ImSig0}) and (\ref{eqn:ImSigV})).
Then, ${\rm Im}~\Sigma^R( \bfk,\omega )$ is calculated using
the usual three-momentum cutoff $\Lambda$.
The real part is then evaluated from   ${\rm Im}~\Sigma^R( \bfk,\omega )$ 
through the dispersion relation.
To calculate ${\rm Re}~\Sigma^R( \bfk,\omega )$ 
with the dispersion relation, we again need to introduce a cutoff.
Here, we assume that the cutoff for the dispersion relation
is the same as the momentum cutoff $\Lambda$, and
we then have 
\begin{eqnarray}
\mbox{Re} \Sigma^R ( \bfk,\omega )
= -\frac1\pi {\rm P} \int_{-\Lambda}^\Lambda d\omega'
\frac{ \mbox{Im}\Sigma^R(\bfk,\omega') }{\omega-\omega'}.
\label{eqn:DispRel2}
\end{eqnarray}
We have checked numerically that the quark spectral function and 
the DOS near the Fermi energy are hardly affected 
by the choice of the cutoff in Eq.~(\ref{eqn:DispRel2}).

\subsection{The quasiparticle picture of quarks}

Now we present the numerical results and
discuss the properties of the quarks above $T_c$.
To investigate the quasiparticle picture of the quarks,
it is useful to consider the behavior of 
 the dispersion relation $ \omega = \omega_\mp( \bfk ) $
and the spectral function $\rho_0( \bfk,\omega )$.
Here,
$\omega_\mp( \bfk ) $  is defined above; see Eq.(\ref{eqn:q-disp}).
A remark is in order here.
Because $\omega_\mp( \bfk ) $ is only the solution to the 
{\em real part} but not the whole part of 
$R_{\mp}( \bfk,\omega_{\mp})$,
$\omega_\mp( \bfk ) $ may not correspond to the peak position of the
spectral function and does not represent physical
excitations when the imaginary part of the Green function is large.

\begin{figure}[btp]
\begin{center}
\begin{tabular}{cc}
\includegraphics[scale=.7]{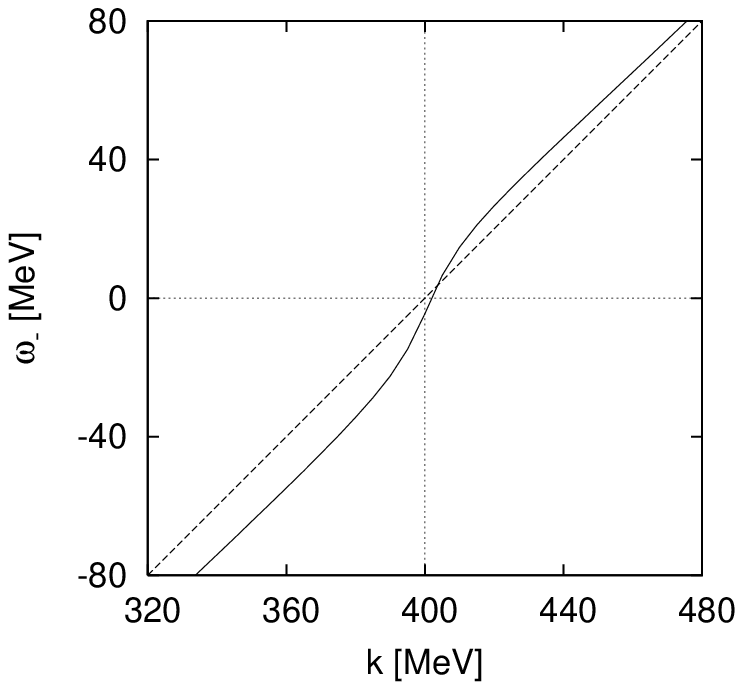} &\hspace{-2cm}
\includegraphics[scale=.7]{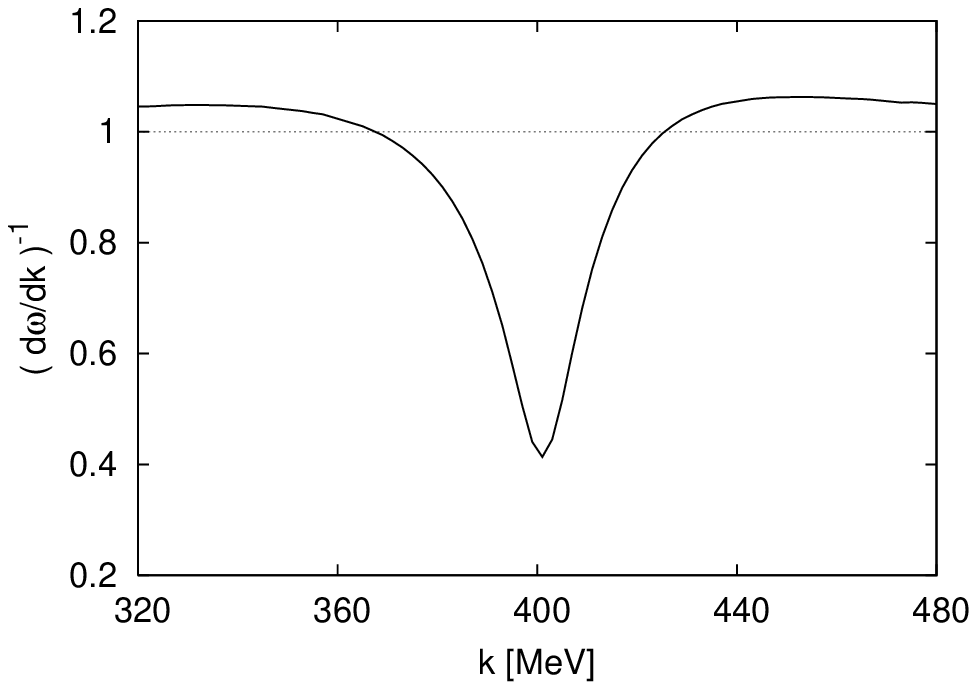}
\end{tabular}
\caption{
The dispersion relation for the quarks,
 $\omega=\omega_-(\bfk)$ in the left panel,
 and 
$ (\partial \omega(\bfk)/\partial |\bfk|)^{-1}$
in the right panel 
with $\mu=400$ MeV and $ \varepsilon \equiv ( T-T_c )/T_c =0.01 $.
Near the Fermi momentum $k_F = \mu$,  
we see a rapid increase of $\omega_-(\bfk)$ and
a pronounced minimum of 
$ (\partial \omega(\bfk)/\partial |\bfk|)^{-1}$ around $k=k_F$,
which means that a gap-like structure is induced 
in the dispersion relation as a precursor to  CSC.
}
\label{fig:dispr}
\end{center} 
\end{figure}

We plot the the dispersion relation of 
the positive energy quarks $\omega=\omega_-(\bfk)$ 
for $\mu=400$ MeV and $\varepsilon=0.01$ 
in the left panel of Fig.~\ref{fig:dispr}.
One sees from the figure that the dispersion relation
deviates from that for free quarks, $\omega=k-\mu$, (dotted line)
and exhibits a rapid increase around the Fermi energy, $ \omega=0$.
We also show $ (\partial \omega(\bfk)/\partial |\bfk|)^{-1}$ 
in the right panel of Fig.~\ref{fig:dispr}.
This quantity is proportional to the DOS,
provided that the imaginary part of the self-energy is ignored.
As shown in the figure,
there appears a pronounced minimum of 
$ (\partial \omega(\bfk)/\partial |\bfk|)^{-1}$ 
near the Fermi momentum, $k_F = \mu$,
reflecting the rapid increase of $\omega_-(\bfk)$.
Although this rapid increase may not be physical because there
exists a large imaginary part of the self-energy in this 
kinematical region,
it may suggest that the gap-like structure is induced 
in the dispersion relation as a precursor to CSC. We will see
that this is the case from the behavior of the spectral function.

\begin{figure}[tbp]
\begin{center}
\includegraphics[scale=0.7]{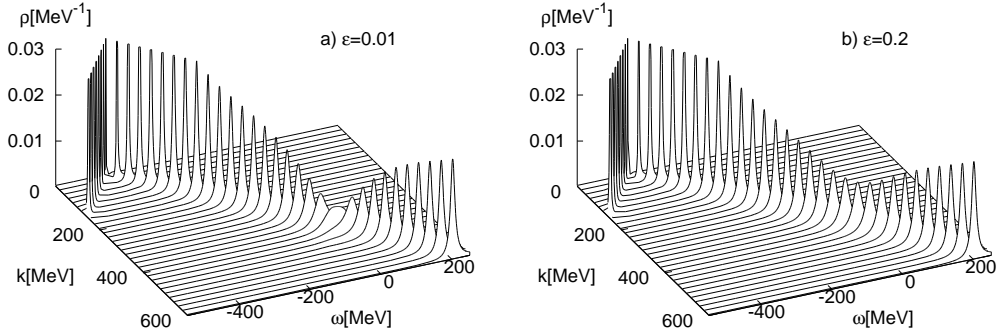}
\caption{
The spectral function $\rho_0$ at $\mu=400$ MeV
and $ \varepsilon=0.01 $ and $0.2$.
The peaks near $ \omega=k-\mu $ and $ \omega=-k-\mu$ correspond
to the quark and anti-quark quasiparticles, respectively.
Note that
there is a depression around $ \omega=0 $,
which is responsible for the pseudogap formation.
}
\label{fig:spc}
\end{center} 
\end{figure}

We show the spectral function 
$\rho_0( \bfk,\omega )$ for $\mu=400$ MeV 
with $\varepsilon=0.01$ and $0.2$ in Fig.~\ref{fig:spc}.
One can see two families of peaks in both figures, near
$ \omega = k-\mu $ and $ \omega = -k-\mu $,
which correspond to the quasiparticle peaks of
the quarks and anti-quarks, respectively.
A notable point is that the quasiparticle peak has a clear depression
around the Fermi energy, $ \omega=0$,
which represents the enhancement of  the decay rate of the quasiparticles
around the Fermi energy.
Thus the depression becomes more pronounced
as $\varepsilon$ becomes smaller.
This behavior is quite different from that of
the conventional Fermi liquids,  
for which the lifetime of the quasiparticles becomes longer
as $\omega$ approaches the Fermi energy, $\omega=0$.

To understand these types of  non-Fermi liquid behavior 
of $\omega_-(\bfk)$ and $\rho_0( \bfk,\omega )$  near $\omega=0$,
we investigate the quark self-energy $\Sigma^R( \bfk,\omega )$.
Before giving the numerical results,
we note that the spectral function of the negative energy
$\rho_+( \bfk,\omega )$ always takes small values around $\omega=0$,
because the real part of $ R_+( \bfk,\omega )$, 
${\rm Re}~R_+( \bfk,\omega ) \simeq \omega + \mu + k$,
does not become smaller around $\omega = 0$.
Thus, $\rho_0( \bfk,\omega )$ near the Fermi energy is 
reproduced by $\rho_-( \bfk,\omega )$ alone:
\begin{eqnarray}
\rho_0( \bfk,\omega ) 
= \frac{ \rho_-( \bfk,\omega ) + \rho_+( \bfk,\omega ) }2
\simeq \frac 12 \rho_-( \bfk,\omega ).
\end{eqnarray}
Therefore, it is sufficient to consider 
the self-energy of the positive energy $\Sigma_-( \bfk,\omega )$
in order to understand the non-Fermi liquid behavior
of $\omega_-(\bfk)$ and  $\rho_0( \bfk,\omega )$.

\begin{figure}[btp]
\begin{center}
\includegraphics[scale=1.3]{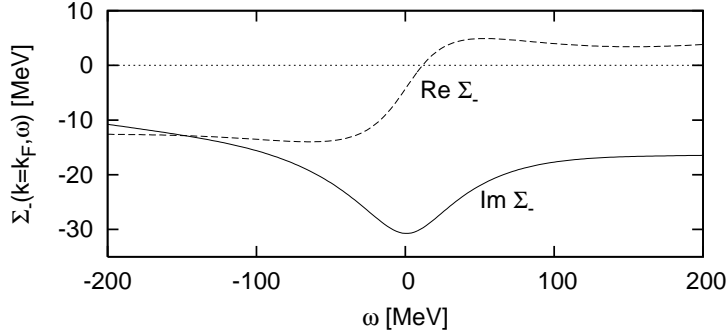} 
\caption{
The self-energy $\Sigma_-$ 
for $ k=k_F$
with $\mu=400$ MeV and $ \varepsilon = 0.01 $.
We observe a peak in Im$\Sigma_-$ 
and a rapid increase of Re$\Sigma_-$ at the same $\omega=0$.
}
\label{fig:sig_kF}
\end{center} 
\end{figure}

In Fig.~\ref{fig:sig_kF}, we plot the real and imaginary parts of 
$\Sigma_-( \bfk,\omega )\vert_{k=k_F}$ for 
 $\mu=400$ MeV and $\varepsilon =0.01$.
One finds 
a peak structure in $|{\rm Im}~\Sigma_-( \bfk,\omega )|$
and a rapid increase in ${\rm Re}~\Sigma_-( \bfk,\omega )$  near $\omega=0$.
These phenomena are closely related to the non-Fermi liquid behavior
of $\rho_0( \bfk,\omega )$ and $\omega_-(\bfk)$. In fact,
the former 
means that the decay rate of the quasiparticles is enhanced
around the Fermi energy, and 
the latter implies an increase of $\omega_-(k)$ around $\omega=0$.

\begin{figure}[bt]
\begin{center}
\includegraphics[scale=.9]{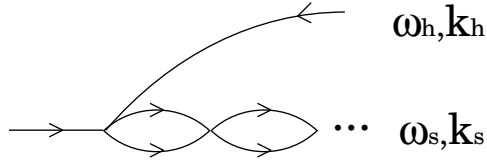}
\caption{
The decay processes of the quarks described in the non-selfconsistent
T-matrix approximation.
}
\label{fig:decay_q}
\end{center} 
\end{figure}

\begin{figure}[bt]
\begin{center}
\vspace{-1cm}
\begin{tabular}{cc}
\hspace{-.8cm}
\includegraphics[scale=.9]{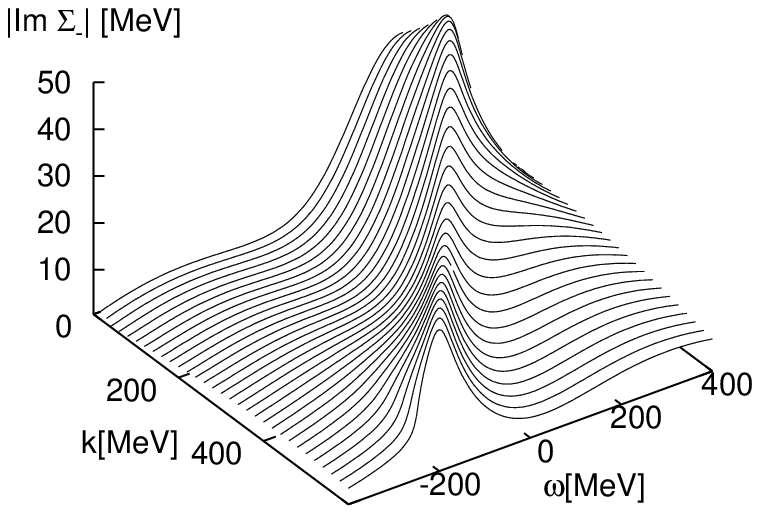} &
\hspace{-1.3cm}
\includegraphics[scale=.9]{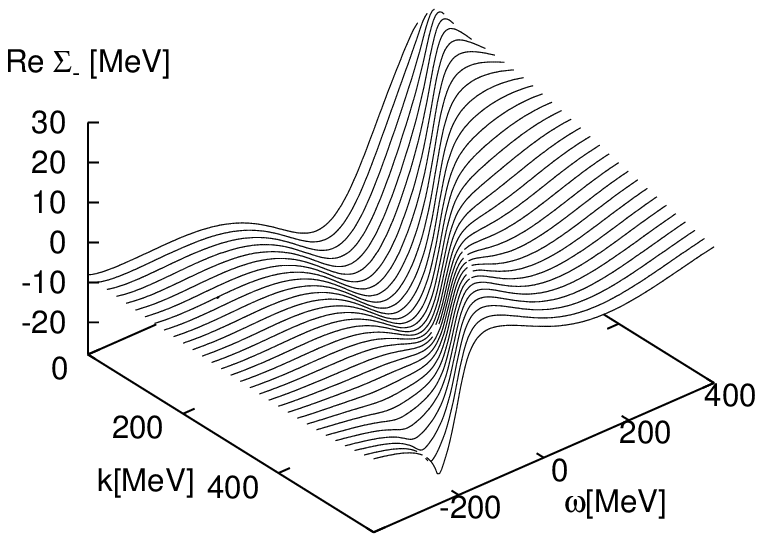}
\end{tabular}
\caption{
The imaginary and real parts of the quark self-energy 
$\Sigma_- (\bfk,\omega) $ with $\mu=400$ MeV 
and $ \varepsilon =0.01 $.
One observes a peak in $| {\rm Im}\Sigma_- |$ 
and a rapid increase of Re$\Sigma_-$ near
$\omega=\mu-k$.
}
\label{fig:sig3d_i}
\end{center} 
\end{figure}

To understand the origin of the characteristic behavior
of $\Sigma_-( \bfk,\omega )$ around $\omega=0$,
let us consider the decay mechanism of a quark.
In our calculation, a quark decays only through the  process
depicted in Fig.~\ref{fig:decay_q},
where an incident quark takes up another quark to make a hole ``h'' and 
a diquark ``(qq)'' state;
\begin{eqnarray}
\mbox q \to {\rm h+(qq)}.
\label{eqn:decaying}
\end{eqnarray}
This process is enhanced
when the diquark pair is in a collective state,
provided that 
the energy-momentum matching is satisfied, i.e.
\begin{eqnarray}
( \omega, \bfk ) = ( \omega_h,\bfk_h ) + ( \omega_s,\bfk_s ),
\label{eqn:DecayProcess}
\end{eqnarray}
where  $(\omega, \bfk)$,
$(\omega_h,\bfk_h)$ and $(\omega_s, \bfk_s)$ denote
the energy-momentum of the decaying particle, 
the hole, and the collective mode composed of the diquark
pair, respectively.
We have seen in \S~3 that the diquark pair field makes
a collective soft mode and the pair fluctuations are strong
near $ \omega=k=0 $ as $T$ approaches $T_c$.
Thus one sees that the decay process Eq.~(\ref{eqn:decaying}) is enhanced 
 for $ ( \omega_s,\bfk_s ) \simeq ( 0,{\bf0} ) $ 
when $T$ is close to $T_c$.
On the other hand, the hole energy should become 
$ \omega_h = \mu-|\bfk_h|$,
since the hole in Eq.~(\ref{eqn:decaying}) is on-shell.
Combining  these two conditions, 
we find that the decay process Eq.~(\ref{eqn:decaying}) is most enhanced
when 
\begin{eqnarray}
( \omega, \bfk ) 
\simeq ( \omega_h,\bfk_h ) = ( \mu - |\bfk|,\bfk ),
\label{eqn:DecayProcess2}
\end{eqnarray}
where we have used the momentum conservation in the last equality.
We  show that this is indeed the case
in the left panel of Fig.~\ref{fig:sig3d_i}, where we show 
the energy-momentum dependence of ${\rm Im}~\Sigma_-( \bfk,\omega )$
for $\mu=400$ MeV and $\varepsilon=0.01$.
There we can clearly see peaks along the line
 $\omega=\mu-|\bfk|$, as expected.
At the Fermi momentum $k=k_F$, this peak corresponds to $\omega=0$,
as shown in Fig.~\ref{fig:sig_kF}.
We also show 
in the right panel of Fig.~\ref{fig:sig3d_i}
the energy-momentum dependence of
${\rm Re}~\Sigma_-( \bfk,\omega )$ for the same $\mu$ and $\varepsilon$.
There a rapid increase in ${\rm Re}~\Sigma_-( \bfk,\omega )$ 
is seen along the line $\omega=\mu-|\bfk|$, at which
 $|{\rm Im}~\Sigma_-( \bfk,\omega )|$ has a peak structure,
in accordance with the behavior expected from
the dispersion relation Eq.~(\ref{eqn:DispRel2}).

\begin{figure}[t]
\begin{center}
\includegraphics[scale=1.4]{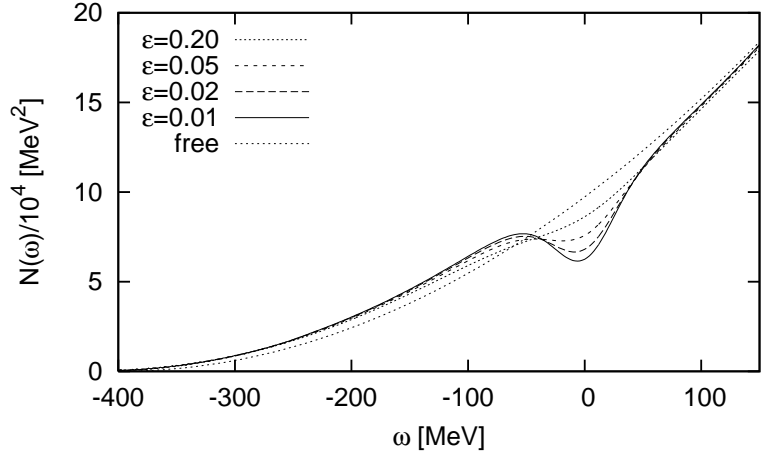} 
\caption{
The density of states for $\mu=400$ MeV and various $\varepsilon$.
The thin dotted curve represents that of free quarks.
A clear pseudogap structure is seen, which survives
up to $ \varepsilon\approx 0.05$.
}
\label{fig:dos}
\end{center} 
\end{figure}

Substituting $\rho_0$ into Eq. (\ref{eqn:N_0}), 
we obtain the DOS, $N(\omega)$, of the quarks 
including the effect of the pair fluctuations.
In Fig.~\ref{fig:dos}, we show the DOS at $\mu = 400$ MeV for 
$\varepsilon = 0.01$, $0.02$, $0.05$ and $0.2$.
The DOS of the free quarks 
\begin{eqnarray}
N_0(\omega) = \frac{ 2 N_f N_c }{ \pi^2 } (\omega-\mu)^2,
\label{eqn:DOS_free}
\end{eqnarray}
is also shown by the thin dotted curves for comparison.
We find that 
there appears a clear depression in the DOS around the Fermi energy
for $\varepsilon=0.01$\cite{ref:KKKN3}.
This is the pseudogap of CSC.
This pseudogap survives up to $\varepsilon \approx 0.05$.
Therefore, the pseudogap certainly does 
appear as an effect of 
the pair fluctuations.
The appearance of the pseudogap in the quark DOS can be naturally
understood as a manifestation of the non-Fermi liquid behavior 
in $\rho_0( \bfk,\omega )$ and $\omega=\omega_-(\bfk)$,
because both the pronounced minimum of the quasiparticle peak and 
the rapid increase of the dispersion relation around $\omega=0$ 
indicate a decrease of the DOS near $\omega=0$.
Although we did not calculate the DOS for smaller $\varepsilon$, say,
$\varepsilon<0.01,$
because of  numerical difficulties,
it seems that the gap-like structure grows as $\varepsilon$ is lowered.
In such a region, however,
the non-linear effects of the pair fluctuations
become significant,
and a systematic incorporation of theses nonlinear effects
using, for instance, the renormalization group analysis,
should be made to  obtain the correct behavior of the DOS.
Such a treatment is, however, beyond the scope of the present work.

\begin{figure}[t]
\begin{center}
\includegraphics[width=14cm]{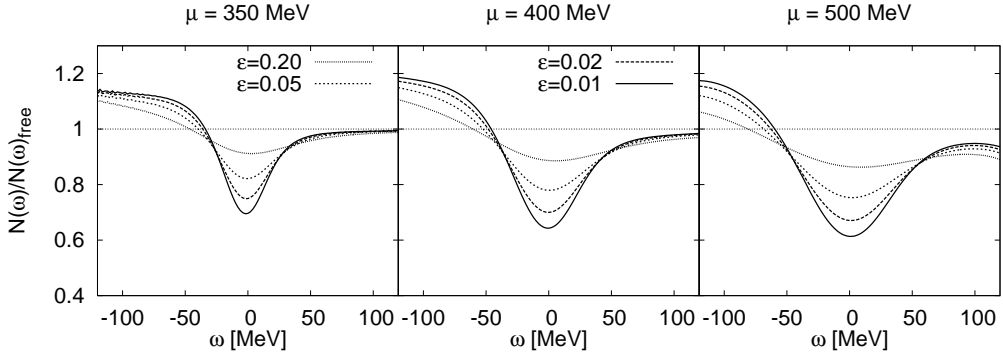} 
\caption{
The chemical potential dependence of the DOS.
Here, the quantities plotted is the ratio of the DOS
for the system under study to that for
 free quarks, $N_0(\omega)$.
The pseudogap becomes more pronounced as $\mu$ increases.
}
\label{fig:dos2}
\end{center} 
\end{figure}

To elucidate the chemical potential dependence of the DOS, 
we show the DOS for $\mu=350,400,500$ MeV in Fig.~\ref{fig:dos2}, where
the values actually plotted are the ratios $N(\omega)/N_0(\omega)$ of 
the DOS for the system under study to that 
of free quarks given in Eq.~(\ref{eqn:DOS_free}).
From Fig.~\ref{fig:dos2}, 
we  can conclude that there is a ``pseudogap region''
within the QGP phase above $T_c$ up to
$T^*=(1.05$ - $1.1)T_c$ at intermediate densities.
We also find that the pseudogap becomes more significant
as $\mu$ increases.
This is because the Cooper instability becomes stronger
as $\mu$ increases owing to the larger Fermi surface.

A comment is in order.
In Ref.~\citen{ref:SRS99}, the pseudogap in low density 
nuclear matter is investigated, and 
it is found that a clear pseudogap is seen 
up to $ \varepsilon\approx0.0025 $,
which is more than one order of magnitude smaller
(in units of $\varepsilon$) than our result.
This is simply a reflection of the
strong coupling nature of QCD in the intermediate density region.
Our result obtained for a three-dimensional 
system reveals that
a considerable pseudogap can be 
formed without 
a low-dimensionality effect, as in the HTSC, and
that the pseudogap phenomena 
may be  universal  in strong coupling
superconductivity\cite{ref:JML97}.

\section{Summary and concluding remarks}

In the present paper,
we have examined precursory fluctuations of the diquark-pair field 
 above the critical temperature $T_c$ for
 two-flavor color superconductivity (CSC) in heated 
quark matter at moderate density.
 A  Nambu-Jona-Lasinio-type model was adopted 
as an effective theory embodying the strong coupling nature
of QCD, which becomes more 
relevant for quark matter at such a relatively low density.
The detailed formulation given here is based on linear-response theory,
and  a detailed account of the calculational procedure
has also been given in the imaginary-time formalism.
We showed that 
the pair fluctuations develop a collective mode whose complex frequency
approaches the origin of the complex energy plane
as $T$ decreases toward $T_c$; i.e., the pairing fluctuations
 form the soft mode of CSC.
Moreover, it was shown that the pairing fluctuations 
are quite strong even well above $T_c$,
owing to the strong-coupling nature of the dynamics,
in comparison with usual metal superconductors.
We have presented an extensive investigation 
 of the properties of the precursory pair field
with finite momenta.
We calculated the spectral function of the pair fluctuations
$\rho(\bfk,\omega)$ as a function of the  momentum $\bfk$ and
energy $\omega$ to clarify the spatial and temporal behavior of 
the pair fluctuations. We also
presented the behavior of the dynamical structure factor.
It was found that the collective mode with
a shorter wavelength has a larger decay width and smaller strength.

We examined the effects of the precursory pair fluctuations on  
the specific heat and the quark spectrum
at values of $T$ close to but above $T_c$.
It was shown that 
$c_{\rm v}$ increases  rapidly when $T$ is lowered toward $T_c$, and
it diverges at $T=T_c$,  along with  singular growth 
of the pair fluctuations, in accordance with 
the general theory of second-order phase transitions. 

The single-quark Green function was also calculated 
 for the first time by incorporating the effects of the 
diquark-pair fluctuations 
in the T-matrix approximation.
We showed that 
the quarks behave like a typical non-Fermi liquid,
owing to the soft mode. In particular,
 the quarks have a larger width near the 
Fermi surface as a result of the interaction with the pairing soft mode.
This leads to  a pseudogap  
in the density of states (DOS) of the quarks in the vicinity of the critical 
point.
The chemical potential  dependence 
of the quark spectrum and the DOS were also studied.
Our results suggest  that the heated quark matter in the vicinity 
of CSC  phase transition at moderate  densities is not 
a simple Fermi liquid, due to the strong fluctuations of the 
pair field. This may invalidate the mean-field approximation.
It is also to be noted that
a pseudogap can form even in (relativistic) three-dimensional systems 
without the need for low-dimensionality, as has been
 suggested for HTSC.
 This leads us to conjecture that 
 the formation of a pseudogap may be a universal phenomenon 
for strongly correlated matter
irrespective of the spatial dimensionality of the system.

\begin{figure}[t]
\begin{center}
\includegraphics[width=10cm]{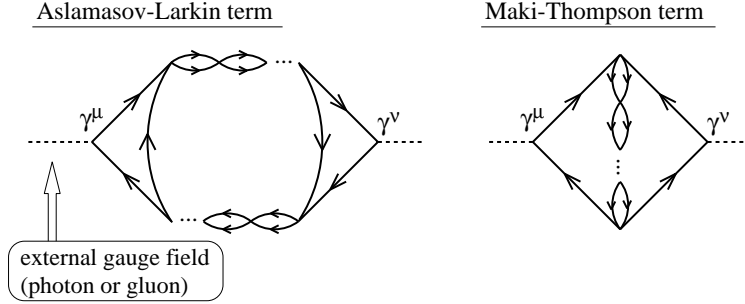} 
\caption{
The diagrams that contribute to 
the photon (or gluon) self-energy,
representing 
the Aslamazov-Larkin term (left) and the Maki-Thompson term (right).
}
\label{fig:AL}
\end{center} 
\end{figure}

What observables can be most strongly 
affected by the precursory fluctuations
of the diquark-pair field? 
It is known in the context of the condensed matter physics 
that pair fluctuations above $T_c$ affect
several transport coefficients,
 including the electric conductivity (EC) and the phonon absorption 
coefficient, as well as the specific heat\cite{ref:AL}.
A large excess of 
electric conductivity seen near but above $T_c$ is known as 
 paraconductivity\cite{ref:AL}. They are attributed to  
pairing fluctuations in the normal phase:
Typical microscopic mechanisms that can
give rise to such an anomalously large conductivity
are those resulting from the so-called Aslamazov-Larkin  and 
Maki-Thompson terms\cite{ref:MT}.
Those are depicted in Fig.~\ref{fig:AL}.
The dotted lines in the figure denote the gauge field, i.e.,
the electro-magnetic field (or the photon) in this case. 
A direct analogy of electric superconductivity to CSC
may be realized by replacing the photon field by the gluon field
in Fig.~\ref{fig:AL}, leading to 
 {\em color-paraconductivity} above $T_c$ in our case. 
However, unfortunately it would  be difficult detect
the color-conductivity directly in experiment or observation.
Nevertheless, note, however, that 
the external photon field can couple to 
the diquark-pair fluctuations of CSC. This coupling is precisely
depicted by the same diagram, Fig.~\ref{fig:AL},
 with the pair field
interpreted as being composed of colored quarks;
in other words, the {\em photon} self-energy in quark matter
at $T>T_c$ is strongly
 modified due to the fluctuating diquark pair field. 
This is interesting, because
modifications of the photon self-energy in 
heated quark matter near $T_c$
should bring about an enhancement of
dilepton emission from the system, which 
may  carry some information concerning
the fluctuations of the diquark-pair field
in heated quark matter, created, say,  in the intermediate stage 
of  heavy-ion collisions.
It is also conceivable that the {\em electric}
conductivity in quark matter
will show anomalous enhancement at $T$ near $T_c$
in the normal phase due to fluctuations of the colored pair-field
through the same process.

The precursory phenomena may also affect the cooling process
of  compact stars.
The temperature of  newborn compact stars 
just after a supernovae can exceed $40$ MeV.
Therefore, dense matter in the interior of 
compact stars may undergo CSC phase transition. 
However, it may be the case that
the $\beta$-equilibrium condition is not completely 
satisfied in proto-compact stars just after supernova 
explosion, in contrast to the situation in 
the interiors of cold compact stars,
as mentioned in the Introduction.
Furthermore, a soft mode necessarily appears if the 
phase transition is second order, irrespective of the pattern of 
CSC. Therefore, the results presented in the preceding sections
can apply to the new-born compact stars.
If it is the case, then  the precursory phenomena
could affect  their cooling process.
Indeed, we have shown that the specific heat is enhanced
through the precursory diquark soft mode as $T$ approaches $T_c$ from
above.
Moreover, the neutrino mean-free path 
should be affected and become shortened through the scattering with
the soft mode near $T_c$.

In the present work, we have employed the non-self-consistent T-matrix 
approximation.
As another, more complicated approximation, there is 
the self-consistent T-matrix approximation, in which
all free fermion propagators are replaced by resummed fermion propagators.
However, the self-consistent approach does not necessarily yield 
better approximation.
As shown in Ref.~\citen{ref:Fuj02}, 
the vertex corrections to the self-energy ignored in the self-consistent 
approximation cancel each other,
and only the lowest-order term survives.
As shown in Ref.~\citen{ref:Fuj02} again,
if one takes into account the vertex corrections
as well as the self-consistency  to the self-energy, their contributions
approximately cancel each other, and correspondingly
the results are rather close to those obtained with the 
non-self-consistent approximation.

In this paper,
 we have used a fixed diquark coupling $G_C$
to examine the $T$ and $\mu$ dependences of the quark spectrum
and the DOS.
An investigation of the quark spectrum with a larger
 $G_C$ may provide a deeper understanding
of the underlying physics for the formation of the pseudogap,
for instance, 
in terms of the 
so called resonance scattering\cite{ref:JML97,ref:HTSC2,ref:KKN05}.

{\bf Acknowledgements}\\ 
M. K. is supported by the Japan Society for the Promotion of Science 
(JSPS) for Young Scientists.
T. Kunihiro is supported by a Grant-in-Aid for Scientific Research by
Monbu-Kagaku-sho in Japan(No.\ 14540263).
Y. N. thanks RIKEN, Brookhaven National Laboratory and the U.S.
Department of Energy for providing the facilities essential for
the completion of this work.
This work is supported by the Grant-in-Aid for the 21st Century COE 
"Center for Diversity and Universality in Physics" 
from the Ministry of Education, Culture, Sports, Science 
and Technology (MEXT) of Japan.

\appendix

\section{Calculation of $Q(\bfk,\omega)$}
\label{sec:appQ}

\begin{figure}[tb]
\begin{center}
\includegraphics[width=10cm]{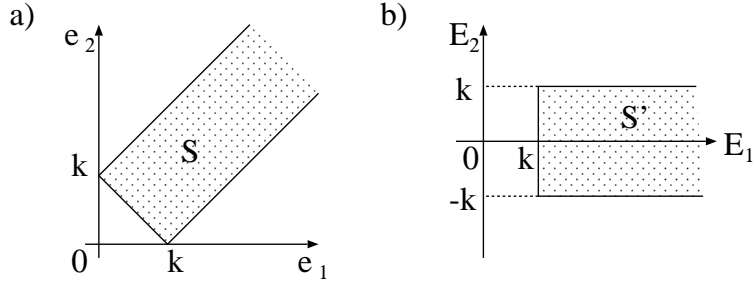}
\caption{The domains of integration in Eqs.~(\ref{eqn:Q_7})
and (\ref{eqn:Q_11}).}
\label{fig:regionS}
\end{center}
\end{figure}

In this appendix, we calculate 
the quark-quark polarization function ${\cal Q}( \bfk ,\nu_n )$ given in 
Eq.~(\ref{eqn:Q}),
which reads
\begin{eqnarray}
{\cal Q}( \bfk ,\nu_n )
&=&
-2 T \sum_m \int \frac{ d^3 \bfp }{ (2\pi)^3 } \mbox{Tr}
\left[
i\gamma_5 \tau_2 \lambda_2 C {\cal G}_0( \bfk-\bfp ,\nu_n-\omega_m )
i\gamma_5 \tau_2 \lambda_2 C {\cal G}_0^T( \bfp ,\omega_m )
\right]
\nonumber \\
&=& -2 N_f ( N_c-1 ) T \sum_m \int \frac{ d^3 \bfp }{ (2\pi)^3 } 
\mbox{Tr}_D \left[ 
{\cal G}_0( \bfk-\bfp ,\nu_n-\omega_m ) {\cal G}_0( \bfp ,\omega_m )
\right],
\label{eqn:Q_1}
\end{eqnarray}
where ${\rm Tr}_D$ denotes the trace over the Dirac spinor.
Here, ${\cal G}_0 (\bfk ,\omega_n )$ is the free Matsubara Green function
\begin{eqnarray}
{\cal G}_0( \bfk ,\omega_n )
&\equiv&
\frac 1{ ( i\omega_n + \mu )\gamma^0 - \bfp \cdot \bfgamma }
= 
\sum_{s=\pm} \frac {\Lambda_s \gamma^0}{ i\omega_n + \mu + s|\bfp| },
\label{eqn:G_s}
\end{eqnarray}
where $\Lambda_\mp = ( 1 \pm \gamma^0 \bfgamma\cdot\hat{\bfk})/2$ 
are the projection operators 
onto positive and negative energies, respectively.
Substituting Eq.~(\ref{eqn:G_s}) into Eq.~(\ref{eqn:Q_1}), we obtain
\begin{eqnarray}
\lefteqn{ {\cal Q}( \bfk ,\nu_n ) } 
\nonumber\\
&=&
-2 N_f ( N_c-1 ) T \sum_m \int \frac{ d^3 \bfp }{ (2\pi)^3 } 
\sum_{s,t=\pm} {\rm Tr}_D
\left[
\frac{ |\bfk-\bfp|\gamma^0 + s( \bfk-\bfp )\cdot\bfgamma }{ 2|\bfk-\bfp| }
\frac{ |\bfp|\gamma^0 + t\bfp\cdot\bfgamma }{ 2|\bfp| }
\right]
\nonumber \\ && \times
\frac1{ i\nu_n - i\omega_m + \mu +s|\bfk-\bfp| }
\frac1{ i\omega_m + \mu + t|\bfp|}
\nonumber \\
&=&
2 N_f ( N_c-1 ) \sum_{s,t=\pm} \int \frac{ d^3 \bfp }{ (2\pi)^3 }
\frac{ |\bfk-\bfp| |\bfp| -st ( \bfk-\bfp ) \cdot \bfp }{ |\bfk-\bfp| |\bfp| }
\frac{ f^+( -s|\bfk-\bfp| ) - f^-( t|\bfp| ) }
{ i\nu_n + 2\mu + s|\bfk-\bfp| + t|\bfp| }
\nonumber \\ 
&=&
N_f ( N_c-1 ) \sum_{s=\pm} \int \frac{ d^3 \bfp }{ (2\pi)^3 }
\left\{
\frac{ ( |\bfk-\bfp|+|\bfp| )^2 -\bfk^2 }{ |\bfk-\bfp| |\bfp| }
\frac{ f^+( -s|\bfk-\bfp| ) - f^-( s|\bfp| ) }
{ i\nu_n + 2\mu + s ( |\bfk-\bfp|+|\bfp| ) }
\right. \nonumber \\ && \left.
-\frac{ ( |\bfk-\bfp|-|\bfp| )^2 -\bfk^2 }{ |\bfk-\bfp| |\bfp| }
\frac{ f^+( -s|\bfk-\bfp| ) - f^-( -s|\bfp| ) }
{ i\nu_n + 2\mu + s ( |\bfk-\bfp|-|\bfp| ) }
\right\}
\label{eqn:Q_A2}
\\
&=&
\frac{ N_f ( N_c-1 ) }{ (2\pi)^2 k }\sum_{s=\pm}
\int_S de_1 de_2
\left\{
\left[ ( e_1+e_2 )^2 -k^2 \right]
\frac{ f^+( -s e_1 ) - f^-( s e_2 ) }
{ i\nu_n + 2\mu - s ( e_1+e_2 ) }
\right. \nonumber \\ && \left.
-\left[ ( e_1-e_2 )^2 -k^2 \right]
\frac{ f^+( -s e_1 ) - f^-( -s e_2 ) }
{ i\nu_n + 2\mu - s ( e_1-e_2 ) }
\right\}.
\label{eqn:Q_7}
\end{eqnarray}
Using Eq.~(\ref{eqn:Q_A2}), Eq.~(\ref{eqn:Q_explicit}) 
can be obtained with a simple manipulation.
In the last equality, we have introduced the new variables 
$e_1 \equiv |\bfk-\bfp| $ and $ e_2 \equiv |\bfp|$, which are
such that
\beq
 \int \frac{ d^3 {\bf p} }{ (2\pi)^3 } 
\frac1{ |{\bf k}-{\bf p}||{\bf p}| }
= \frac1{ (2\pi)^2 k } \int_S de_1 de_2,
\eeq
 with  $S$ being the domain of integration shown in 
Fig.~\ref{fig:regionS} a).

Performing the analytic continuation 
$Q^R( \bfk,\omega) = {\cal Q}( \bfk,\nu_n)|_{i\nu_n=\omega+i\eta}$
and using the fact that the domain $S$ is invariant
under the exchange of $e_1$ and $e_2$, we have
\begin{eqnarray}
Q^R ( \bfk , \omega )
&=& 
\frac{ N_f (N_c-1) }{ (2\pi)^2 k } \int_S de_1 de_2
\nonumber \\ &&
\times \left\{
[ ( e_1 + e_2 )^2 - k^2 ] \left(
\frac{ 1 - 2f^-( e_1 ) }{ \omega + 2\mu - e_1 - e_2 + i\eta }
-\frac{ 1 - 2f^+( e_1 ) }{ \omega + 2\mu + e_1 + e_2 + i\eta } 
\right) \right.\nonumber \\ 
&& \left.
+2[ ( e_1-e_2 )^2 - k^2 ]
\frac{ f^-( e_1 ) - f^+( e_2 ) }{\omega + 2\mu - e_1 + e_2 + i\eta }
\right\}\label{eqn:Q_10}
\\
&=&
\frac{ N_f (N_c-1) }{ (2\pi)^2 k } \int_{S'} \frac{ dE_1 dE_2 }2
\nonumber \\
&&\times \left\{
( E_1^2 - k^2 ) \left(
\frac{ \tanh \frac{ E_1+E_2 - 2\mu }{4T} }
{ \omega + 2\mu - E_1 + i\eta }
-\frac{ \tanh \frac{ E_1+E_2 + 2\mu }{4T} }
{ \omega + 2\mu + E_1 + i\eta } \right) \right. \nonumber \\
&& \left.
+2( E_2^2 - k^2 )
\frac{ f^-( \frac{ E_1+E_2 }2 ) - f^+( \frac{ E_1-E_2 }2 ) }
{\omega + 2\mu - E_2 + i\eta } 
\right\}. \label{eqn:Q_11}
\end{eqnarray}
In the second equality, the new variables
$ E_1 \equiv e_1+e_2 $ and $ E_2 \equiv e_1-e_2 $ have been introduced
for later convenience to obtain the imaginary part of $Q^R(\bfk,\omega)$.

As explained in the main text(\S 3), we 
first calculate the imaginary part ${\cal Q}$
without introducing a cutoff, and then we define the real part 
using  the dispersion relation with a momentum cutoff introduced.
From Eq.~(\ref{eqn:Q_11}), we have for the imaginary part of $Q^R$, 
\begin{eqnarray}
\lefteqn{
{\rm Im} Q^R( \bfk , \omega ) }
\nonumber \\
&=& 
-\pi \frac{ N_f (N_c-1) }{ 8\pi^2 k } \int_k^\infty dE_1 \int_{-k}^k dE_2
\nonumber \\
&&\times \left\{
( E_1^2 - k^2 ) \left( 
\tanh \textstyle \frac{ E_1+E_2 - 2\mu }{4T}
\delta( \omega + 2\mu - E_1 )
-\tanh \textstyle \frac{ E_1+E_2 + 2\mu }{4T}
\delta( \omega + 2\mu + E_1 ) \right) \right. \nonumber \\
&& \left.
+2( E_2^2 - k^2 )
\{ f( \textstyle \frac{ E_1+E_2 }2 -\mu ) 
- f( \textstyle \frac{ E_1-E_2 }2 +\mu ) \}
\delta( \omega + 2\mu - E_2 )
\right\} \nonumber \\
&=&
-\frac{ N_f (N_c-1) }{ 8\pi k } \left\{
\theta( |\omega+2\mu|-k )
\int_{-k}^k dE_2 \left[ (\omega+2\mu)^2 - k^2 \right]
\tanh \textstyle \frac{ \omega+E_2 }{4T}
\right. \nonumber \\
&& \left.
+2 \theta( -|\omega+2\mu|+k )
\int_k^\infty dE_1 \left[ (\omega+2\mu)^2 - k^2 \right]
\{ f( \textstyle\frac{ E_1+\omega }2 ) - f( \textstyle\frac{ E_1-\omega }2 ) \}
\right\} \nonumber  \\
&=&
-\frac{ N_f ( N_c-1 ) T }{ 2\pi k }
\left[ ( \omega+2\mu )^2 - k^2 \right]
\left\{
\theta( |\omega+2\mu| - k )
\log \frac{ \cosh ( \omega+k )/4T }{ \cosh ( \omega-k )/4T }\right.
\nonumber \\
&&
\qquad\qquad\qquad\qquad\qquad\qquad
\left. +\theta( -|\omega+2\mu| + k )
\log \frac{ 1 + \e^{ -( \omega+k )/2T } }
{ 1 + \e^{ -( -\omega+k )/2T } } \right\}
\nonumber \\
&=&
-\frac{ N_f ( N_c-1 ) T }{ 2\pi k }
\left[ ( \omega+2\mu )^2 - k^2 \right]
\nonumber \\ &&\times
\left\{
\log \frac{ \cosh ( \omega+k )/4T }{ \cosh ( \omega-k )/4T }
-\frac \omega{2T} \theta( -|\omega+2\mu| + k ) \right\}.
\label{eqn:ImQ_5}
\end{eqnarray}

As stated above, we next define the real part of $Q^R$
through  the dispersion relation with Eq.~(\ref{eqn:ImQ_5}) for the 
imaginary part:
\begin{eqnarray}
{\rm Re}Q^R(\bfk,\omega) 
= -\frac 1\pi {\rm P}
\int d\omega' \frac{ {\rm Im}Q(\bfk,\omega) }{\omega-\omega'}.
\label{eqn:DispR_app}
\end{eqnarray}
We  introduce a cutoff so that
the integration of Eq.~(\ref{eqn:DispR_app}) 
becomes consistent with  the Thouless criterion Eq.~(\ref{eqn:Thouless}).
With vanishing momentum, Eq.~(\ref{eqn:ImQ_5}) becomes
\begin{eqnarray}
{\rm Im }Q^R( \bfk\to {\bf 0} ,\omega )
= - \frac{ N_f ( N_c-1 ) }{ 4\pi } ( \omega + 2\mu )^2 
\tanh \frac\omega{4T} + O(k^2).
\end{eqnarray}
Substituting this equation into Eq.~(\ref{eqn:DispR_app}), we have
\begin{eqnarray}
{\rm Re}Q^R( {\bf 0},0 ) 
= \frac { N_f ( N_c-1 ) }{ 4\pi^2 } {\rm P} \int d\omega 
\frac{ ( \omega + 2\mu )^2 \tanh \frac\omega{4T} }\omega.
\end{eqnarray}
In comparison with Eq.~(\ref{eqn:GE_Tc}),
we find that the cutoff should be introduced
as  in Eq.~(\ref{eqn:DispRel}).

\section{Analytic Properties of $Q(\bfk,\omega)$}
\label{sec:appQ2}

In order to find the pole of the response function $D^R(\bfk,\omega)$,
we have to derive the analytic form of $Q^R(\bfk,\omega)$
in the lower-half complex energy plane, as given in 
Eq.~(\ref{eqn:QR_lower}).

We here restate the definitions 
of $Q(\bfk,z)$, $Q^R(\bfk,z)$ and $Q^A(\bfk,z)$
in the complex-energy plane $\mathbb{C}$
\begin{eqnarray}
Q( \bfk,z ) \equiv {\cal Q}( \bfk,\nu_n)|_{ i\nu_n \to z }, \\
Q^R( \bfk,z ) \equiv Q^R( \bfk,\omega )|_{ \omega \to z }, \\
Q^A( \bfk,z ) \equiv Q^A( \bfk,\omega )|_{ \omega \to z }.
\label{eqn:Q_defs}
\end{eqnarray}
Because $Q(\bfk,z)$ has a cut along the real axis,
$Q^R(\bfk,z)$ ($Q^A(\bfk,z)$) coincides with $Q(\bfk,z)$
only in the upper (lower) half plane $\mathbb{C}^+$ ($\mathbb{C}^-$):
\begin{eqnarray}
&& Q(k,z) = Q^R(k,z) \qquad \mbox{for} \quad z \in \mathbb{C}^+, \nonumber \\
&& Q(k,z) = Q^A(k,z) \qquad \mbox{for} \quad z \in \mathbb{C}^-.
\label{eqn:Q_RA}
\end{eqnarray}

In this appendix, we wish to find the analytic form of 
$Q^R(\bfk,z)$ in $\mathbb{C}^-$.
To accomplish this,
we first evaluate $Q^R(\bfk,z)$ just below the real axis,
 at $z=\omega-i\eta$, and then determine $Q^R(\bfk,z)$ in $\mathbb{C}^-$
by analytic continuation.
Here, we note that the discontinuity of $Q(\bfk,z)$ at the cut 
along the real axis is characterized by 
the discontinuity of the imaginary part, that is,
\begin{eqnarray}
\mbox{Re} Q^R( k, \omega +i\eta ) 
&=& \mbox{Re} Q( k,\omega +i\eta ) = \mbox{Re} Q( k,\omega -i\eta )
= \mbox{Re} Q^A( k, \omega -i\eta ) , 
\label{eqn:Q_RA_Real} \\
\mbox{Im} Q^R( k, \omega +i\eta ) 
&=& \mbox{Im} Q( k,\omega +i\eta ) = -\mbox{Im} Q( k,\omega -i\eta )
= -\mbox{Im} Q^A( k, \omega -i\eta ) 
\nonumber \\
&\equiv& I(k,\omega),
\label{eqn:Q_RA_Im}
\end{eqnarray}
with $\omega$ being a real variable.
We  find from Eq.~(\ref{eqn:ImQ})
that cuts of $I(\bfk,z)\equiv I(\bfk,\omega)|_{\omega\to z}$
exist at $ \mbox{Im}z = 2(2n+1)\pi T $.
Hence, this function is analytic in $ \mathbb{C}_I \equiv 
\{z \in \mathbb{C} | -2i\pi T < \mbox{Im}z < 2i\pi T \}$.
Using Eqs.(\ref{eqn:Q_RA_Real}) and (\ref{eqn:Q_RA_Im}),
we find
\begin{eqnarray}
Q^R ( k, \omega ) - iI ( k,\omega )
&=&
Q^R ( k, \omega +i\eta ) - i{\rm Im} Q^R ( k, \omega +i\eta )
\nonumber \\
&=&{\rm Re} Q^R( k, \omega +i\eta ) = {\rm Re} Q^A( k, \omega -i\eta )
\nonumber \\
&=& Q^A ( k, \omega -i\eta ) - i{\rm Im} Q^A ( k, \omega -i\eta ),
\nonumber \\
&=& Q ( k, \omega -i\eta ) + iI ( k,\omega ),
\label{eqn:Q_oncut}
\end{eqnarray}
which leads to 
\begin{eqnarray}
Q^R ( k, \omega -i\eta )
&=& Q^R ( k, \omega +i\eta )
\nonumber \\
&=& Q ( k, \omega -i\eta ) +2iI ( k,\omega ).
\label{eqn:QR_lower2}
\end{eqnarray}
In the first equality here, 
we have used the fact that $Q^R( \bfk,z )$ is a continuous function 
on the real axis,
and Eq.~(\ref{eqn:Q_oncut}) is applied to derive the second equality.
Equation (\ref{eqn:QR_lower2}) implies that we can express
$Q^R(\bfk,z)$ just below the real axis 
in terms of the functions $Q(\bfk,z)$ and $I(\bfk,z)$, whose analytic forms
in $\mbox{Im}z <0$ are already known.
Because both $Q(\bfk,z)$ and $I(\bfk,z)$ are analytic functions
in $\mathbb{C}^- \cap \mathbb{C}_I$,
the analytic form of $Q^R$ in 
this domain can be obtained straightforwardly
from the uniqueness theorem of the analytic continuation as 
\begin{eqnarray}
Q^R ( k,z ) = Q ( k,z ) + 2iI ( k,z ), 
\qquad {\rm for} \quad z \in \mathbb{C}^- \cap \mathbb{C}_I.
\label{eqn:QR_lower_app}
\end{eqnarray}
Although Eq.~(\ref{eqn:QR_lower_app}) is valid only for 
${\rm Im}z > -2\pi T$,
this region is sufficient to find the pole.

\begin{figure}[tb]
\begin{center}\leavevmode
\includegraphics{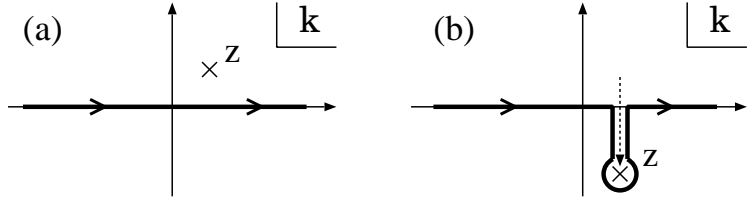} 
\caption{
The analytic continuation of $Q^R(\bfk,\omega)$ 
for a vanishing momentum.
}
\label{fig:path_k0}
\end{center} 
\end{figure}

For vanishing momentum,
analytic continuation into the lower half plane 
is understood as changing the contour of the momentum integration
as shown in Fig.~\ref{fig:path_k0}.
In this case, the term originating from the pole integration 
is added in the lower half plane\cite{ref:KKKN1}, and we obtain
\begin{eqnarray}
Q^R({\bf 0}, z \in \mathbb{C}^- ) &=&
\frac{2N_f(N_c-1)}{\pi^2}\big[
{\rm P}\int_{-\Lambda}^{\Lambda}
k^2 dk\frac{\tanh \frac{|{\bf k}|-\mu}{2T}}{z-2(|\bfk|-\mu)}\nonumber \\
&&-\frac12 2\pi i(\mu+\frac z2)^2\tanh \frac z{4T}
\big].
\label{eqn:QR_lowerk0}
\end{eqnarray}
Equation~(\ref{eqn:QR_lowerk0}) coincides, of course, with
the $k \to 0$ limit of Eq.~(\ref{eqn:QR_lower_app}).

\section{Calculation of $\Sigma$}
\label{sec:appSigma}

\begin{figure}[tbp]
\begin{center}
\includegraphics[scale=.8]{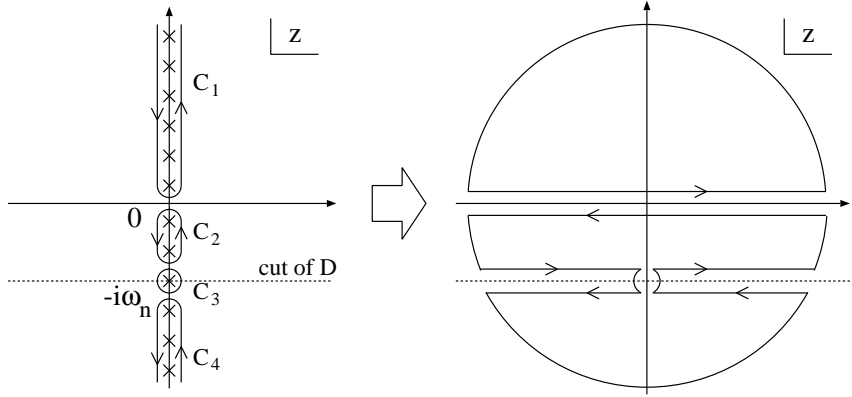} 
\caption{
The contours of the $q_0$ integrals in Eqs.~(\ref{eqn:Sig})
and (\ref{eqn:Sig_2}).
}
\label{fig:contour}
\end{center}
\end{figure}

To carry out the summation of the Matsubara frequency $\omega'_m$ 
in Eq.~(\ref{eqn:SE1}), 
it is sufficient to consider the case  $\omega_n >0$.
Because both $\Xi( \bfk,z )$ and $G_0( \bfk,z )$ have
a cut along the real axis,
the integrand of Eq.~(\ref{eqn:SE1}) has two cuts
along the horizontal lines 
$ \mbox{Im}z = 0 $ and $ \mbox{Im}z = -i\omega_n $
in the complex energy plane as shown in Fig.~\ref{fig:contour}.
Therefore, 
we must to separate the Matsubara frequency into four parts
to carry out the Matsubara sum in Eq.~(\ref{eqn:SE1}):
\begin{eqnarray}
\tilde{\Sigma}( \bfp,\omega_n )
&=& 8T \sum_m \int \frac{d^3 \bfq }{(2\pi)^3}
\tilde\Xi( \bfp +\bfq, \omega_n +\omega_m )
{\cal G}_0( \bfq,\omega_m )
\nonumber \\
&=&
4 \int_{C_1} \frac{dq^0}{2\pi i}
\tanh \frac{q^0}{2T} \int \frac{d^3 \bfq}{(2\pi)^3}
\Xi^R ( \bfp+\bfq, i\omega_n+q^0 ) G_0^R( \bfq,q^0 )  
\nonumber \\
&+&
4 \int_{C_2} \frac{dq^0}{2\pi i}
\tanh \frac{q^0}{2T}\int \frac{d^3 \bfq}{(2\pi)^3}
\Xi^R ( \bfp+\bfq, i\omega_n+q^0) G_0^A ( \bfq,q^0 )  
\nonumber \\
&+&
4 \int_{C_3} \frac{dq^0}{2\pi i}
\tanh \frac{q^0}{2T}\int \frac{d^3 \bfq}{(2\pi)^3}
\Xi( \bfp+\bfq, i\omega_n+q^0)
G_0^A( \bfq,q^0 )  
\nonumber \\
&+&
4 \int_{C_4} \frac{dq^0}{2\pi i}
\tanh \frac{q^0}{2T}\int \frac{d^3 \bfq}{(2\pi)^3}
\Xi^A( \bfp+\bfq, i\omega_n+q^0)
G_0^A ( \bfq,q^0 ),
\label{eqn:Sig}
\end{eqnarray}
where the contours $ C_1$ - $C_4 $ are shown in the left panel of
Fig.~\ref{fig:contour}.
The definitions of the retarded and advanced T-matrices
$\Xi^R$ and $\Xi^A$ are 
\begin{eqnarray}
\Xi( \bfp,z ) &=& \Xi^R( \bfp,z ) 
\quad \mbox{for   } z \in\mathbb{C}^+,
\nonumber \\
\Xi( \bfp,z ) &=& \Xi^A( \bfp,z )
\quad \mbox{for   } z \in\mathbb{C}^-,
\end{eqnarray}

Next, we change the contours $ C_1$ - $ C_4 $,
as shown in Fig.~\ref{fig:contour}.
Then the contribution from the integral along the large circle 
vanishes, and only the integrals along the horizontal lines 
$ \mbox{Im}\omega = 0 $ and $ \mbox{Im}\omega = -i\omega_n $ remain.
Thus, we have 
\begin{eqnarray}
\tilde{\Sigma}( \bfp,\omega_n )
&=& 4 \int \frac{d^3 \bfq}{(2\pi)^3}
\int^{\infty}_{-\infty}\frac{dq^0}{2\pi i} \tanh \frac{q^0}{2T}
\nonumber \\
&&\times
\Xi^R ( \bfp+\bfq, i\omega_n +q^{0})
\left\{ G_0^R( \bfq,q^0 ) - G_0^A( \bfq,q^0 ) \right\} 
\nonumber \\ 
&+& 4 \int \frac{d^3 \bfq}{(2\pi)^3}
\int^{\infty-i\omega_n}_{-\infty-i\omega_n}\frac{dq^0}{2\pi i}
\tanh \frac{q^0}{2T}
\nonumber \\
&&\times \left\{
\Xi^R ( \bfp+\bfq,i\omega_m+q^0 ) -\Xi^A ( \bfp+\bfq,i\omega_m+q^0 )
\right\} G_0^A( \bfq,q^0 )
\nonumber \\
&=& 2 \int \frac{ d^4q }{ (2\pi)^4 }
\left\{ \tanh \frac{ q_0 }{ 2T }
\Xi^R ( \bfp+\bfq,i\omega_n+q^0 ) {\rm Im}G_0^R( \bfq,q^0 )
\right. \nonumber \\ && \left. \qquad \qquad 
+ \tanh \frac{ q^0-i\omega_n }{2T}
{\rm Im}\Xi^R ( \bfp+\bfq,q^0 ) G_0^A( \bfq,q^0-i\omega_n)
\right\}, \qquad 
\label{eqn:Sig_2}
\end{eqnarray}
where $ \int d^4q/(2\pi)^4 
\equiv \int dq^0/(2\pi) \int d^3\bfq/(2\pi)^3 $, 
and the relations
\begin{eqnarray}
\mbox{Im} \Xi^R  = ( \Xi^R - \Xi^A )/2i, \qquad
\mbox{Im} G_0^R  = ( G_0^R - G_0^A )/2i
\end{eqnarray}
have been used in Eq.~(\ref{eqn:Sig_2}).
It is known that the singularity corresponding to the integral 
along $C_3$ does not give rise to any problem\cite{ref:AGD}.
Then, bearing in mind that 
$ \tanh \frac{ \omega-i\omega_n }{2T} = \coth \frac{ \omega }{2T} $,
and employing the analytic continuation
$\Sigma^R({\bfk},\omega) 
= \tilde{\Sigma}(\bfk,\omega_n)|_{i\omega_n=\omega + i\eta}$,
we obtain the self-energy in real time,
\begin{eqnarray}
\Sigma^R ( \bfk,\omega )
&=&
2 \int \frac{ d^4q}{ (2\pi)^4 }
\left\{
\tanh \frac{q^0}{ 2T }
\Xi^R ( \bfp+\bfq,\omega+q^0 ) {\rm Im} G_0^R( \bfq,q^0 )
\right. \nonumber \\ && \left. \qquad \qquad \qquad 
+ \coth \frac{q^0}{2T}
{\rm Im}\Xi^R ( \bfp+\bfq,q^0 ) G_0^A( \bfq,q^0-\omega )
\right\}.
\label{eqn:Sig^Rapp}
\end{eqnarray}

To simplify Eq.~(\ref{eqn:Sig^Rapp}) further,
we substitute the explicit form of the free Green function, 
\begin{eqnarray}
G_0^{\{R,A\}}(q) 
= \frac{ \Lambda_- (\bfq) }{ q^0+\mu \pm i\eta - |\bfq| }
+ \frac{ \Lambda_+ (\bfq) }{ q^0+\mu \pm i\eta + |\bfq| },
\end{eqnarray}
and its imaginary part,
\begin{eqnarray}
\mbox{Im} G_0^{\{R,A\}}(q) 
= \mp \pi \left(
  \Lambda_- (\bfq) \delta( q^0+\mu - |\bfq| )
+ \Lambda_+ (\bfq) \delta( q^0+\mu + |\bfq| ) \right),
\end{eqnarray}
into Eq.~(\ref{eqn:Sig^Rapp}).
Then, Eq.~(\ref{eqn:Sig^Rapp}) becomes 
\begin{eqnarray}
\Sigma_0 ( \bfp,\omega ) 
&=& \frac14 \Tr\{\gamma^0 \Sigma ( \bfp,\omega )\} \nonumber \\
&=& 
-\frac12
\int \frac{d^3 \bfq}{(2\pi)^3}\int \frac{d\omega}{2\pi}
\frac{{\rm Im}~\Xi^R ( \bfp+\bfq,\omega)
}{\omega-p^0-|\bfq|+\mu-i\eta}
[
\tanh \frac{|\bfq|-\mu}{2T}
- \coth\frac{\omega}{2T}
] \nonumber \\
&&-\frac{1}{2}\int \frac{d^3 \bfq}{(2\pi)^3}\int \frac{d\omega}{2\pi}
\frac{{\rm Im}~\Xi^R ( \bfp+\bfq,\omega )
}{\omega-p^0+|\bfq|+\mu-i\eta}
[
\tanh \frac{-|\bfq|-\mu}{2T}
- \coth\frac{\omega}{2T}
], \nonumber \\
\\
\Sigma_{\rm v}( \bfp,\omega )
&=& -\frac14\Tr\{(\hat{\bfp}\cdot \bfgamma) \Sigma ( \bfp,\omega ) \} 
\nonumber \\
&=&  
-\int \frac{d^3 \bfq}{(2\pi)^3}\int \frac{d\omega}{2\pi}
\frac{{\rm Im}~\Xi^R ( \bfp+\bfq,\omega)
}{\omega-p^0-|\bfq|+\mu-i\eta}
( \hat{\bfq} \cdot \hat{\bfp} ) 
[\tanh \frac{|\bfq|-\mu}{2T}- \coth\frac{\omega}{2T}] 
\nonumber \\
&&
+\int \frac{d^3 \bfq}{(2\pi)^3}\int \frac{d\omega}{2\pi}
\frac{{\rm Im}~\Xi^R ( \bfp+\bfq,\omega )
}{\omega-p^0+|\bfq|+\mu-i\eta}
( \hat{\bfq} \cdot \hat{\bfp} )
[ \tanh \frac{-|\bfq|-\mu}{2T} - \coth\frac{\omega}{2T}].
\nonumber \\
\end{eqnarray}
The imaginary parts of these quantities are found to be
\begin{eqnarray}
\lefteqn{ {\rm Im}~\Sigma_0( \bfp,p^0 ) }
\nonumber \\
&=& 
-\frac14
\int \frac{d^3 \bfq}{(2\pi)^3}
{\rm Im}~\Xi^R( \bfp+\bfq, p^0 + |\bfq|-\mu )
[ \tanh \textstyle \frac{|{\bfq}|-\mu}{2T}
- \coth\frac{p^0 + |{\bfq}|-\mu}{2T}
] \nonumber \\
&&-\frac14 \int \frac{d^3 \bfq}{(2\pi)^3}
{\rm Im}~\Xi^R( \bfp+\bfq, p^0-|\bfq|-\mu )
[ \tanh \textstyle \frac{-|{\bfq}|-\mu}{2T} 
- \coth\frac{p^0 -|{\bfq}| -\mu}{2T} ], 
\label{eqn:ImSig0}
\\
\lefteqn{ {\rm Im}~\Sigma_{\rm v}(\bfp,p^0) }
\nonumber \\
&=& 
-\frac14 \int \frac{d^3 \bfq}{(2\pi)^3}
{\rm Im}~\Xi^R ( \bfp+\bfq, p^0+|\bfq|-\mu)
( \hat{\bfq} \cdot \hat{\bfp} )
[
\tanh \textstyle \frac{|{\bfq}|-\mu}{2T}
- \coth \textstyle \frac{p^0+|{\bf q}|-\mu}{2T}
] \nonumber \\
&&
+\frac14 \int \frac{d^3 \bfq}{(2\pi)^3}
{\rm Im}~\Xi^R ( \bfp+\bfq, p^0 - |\bfp| -\mu)
( \hat{\bfq} \cdot \hat{\bfp} )
[
\tanh \textstyle \frac{-|{\bfq}|-\mu}{2T}
- \coth \textstyle \frac{p^0 - |{\bfp}| -\mu}{2T}
].\nonumber \\
\label{eqn:ImSigV}
\end{eqnarray}

\end{document}